\documentclass[twoside]{IEEEtran}
\IEEEoverridecommandlockouts
\normalsize

\def\Figs{./} 
\usepackage[cmex10]{amsmath} 

\usepackage[noadjust]{cite}       

\usepackage{amsmath}
\usepackage{amsopn}
\usepackage{amssymb}
\usepackage{accents}
\usepackage{mathrsfs}
\usepackage{pifont}

\usepackage{bm}         
\usepackage{fix-cm}     

\usepackage{mleftright} 
\mleftright                 

\usepackage{mathdots}   

\newcommand{\mystrut}{\IEEEeqnarraystrutmode\IEEEeqnarraystrutsizeadd{2pt}{1pt}}
\makeatletter
\def\IEEElabelanchoreqn#1{\bgroup
\def\@currentlabel{\p@equation\theequation}\relax
\def\@currentHref{\@IEEEtheHrefequation}\label{#1}\relax
\Hy@raisedlink{\hyper@anchorstart{\@currentHref}}\relax
\Hy@raisedlink{\hyper@anchorend}\egroup}
\makeatother

\usepackage{graphicx}
\usepackage{enumitem}

\usepackage[utf8]{inputenc} 
\usepackage[T1]{fontenc} 

\usepackage{hyphenat}

\newcommand{\midk}[1]{\kern0.1em #1 \kern0.1em}
\newcommand{\middlek}[1]{\kern0.1em \middle#1 \kern0.1em}
\newcommand{\bigk}[1]{\kern-0.1em \bigm#1 \kern-0.1em}
\newcommand{\Bigk}[1]{\kern-0.1em \Bigm#1 \kern-0.1em}
\newcommand{\biggk}[1]{\kern-0.1em \biggm#1 \kern-0.1em}
\newcommand{\Biggk}[1]{\kern-0.1em \Biggm#1 \kern-0.1em}

\newcommand{\vect}[1]{\mathbf{#1}} 
\newcommand{\vectg}[1]{\bm{#1}} 
\renewcommand{\vect}[1]{\vectg{#1}} 

\newcommand{\mat}[1]{\mathsf{#1}} 
 
\newcommand{\inv}[1]{#1^{-1}} 
\newcommand{\trans}[1]{#1^{\textup{\textsf{\tiny T}}}} 

\newcommand{\Reals}{\mathbb{R}}      
\newcommand{\Naturals}{\mathbb{N}}   

\newcommand{\set}[1]{\mathcal{#1}}           
\newcommand{\card}[1]{\left|#1\right|}       

\newcommand{\bigcard}[1]{\bigl|#1\bigr|}

\newcommand{\const}[1]{\textnormal{\usefont{U}{eur}{m}{n}\selectfont #1}} 

\newcommand{\HH}{\mathop{}\!\const{H}}  
\newcommand{\II}{\mathop{}\!\const{I}}  

\newcommand{\HP}[1]{\HH\left(#1\right)} 
\newcommand{\eHP}[1]{\HH(#1)} 
\newcommand{\bigHP}[1]{\HH\bigl(#1\bigr)}

\newcommand{\HPcond}[2]{\HH\left(#1 \kern0.1em\middle|\kern0.1em #2\right)}
\newcommand{\eHPcond}[2]{\HH(#1 \kern0.1em|\kern0.1em #2)} 
\newcommand{\bigHPcond}[2]{\HH\bigl(#1 \kern-0.1em \bigm| \kern-0.1em#2\bigr)}
\newcommand{\BigHPcond}[2]{\HH\Bigl(#1 \kern-0.1em \Bigm| \kern-0.1em#2\Bigr)}

\newcommand{\MinH}{\HH_\infty}  
\newcommand{\MH}[1]{\MinH\left(#1\right)} 
\newcommand{\eMH}[1]{\MinH(#1)}

\newcommand{\MHcond}[2]{\MinH\left(#1 \kern0.1em\middle|\kern0.1em #2\right)}
\newcommand{\eMHcond}[2]{\MinH(#1 \kern0.1em|\kern0.1em #2)} 
\newcommand{\bigMHcond}[2]{\MinH\bigl(#1 \kern-0.1em \bigm| \kern-0.1em#2\bigr)}
\newcommand{\BigMHcond}[2]{\MinH\Bigl(#1 \kern-0.1em \Bigm| \kern-0.1em#2\Bigr)}

\newcommand{\MI}[2]{\II\left(#1 \kern0.1em{;}\kern0.1em #2\right)} 
\newcommand{\eMI}[2]{\II(#1 \kern0.1em{;}\kern0.1em #2)} 
\newcommand{\bigMI}[2]{\II\bigl(#1 \kern0.1em{;}\kern0.1em #2\bigr)}
\newcommand{\BigMI}[2]{\II\Bigl(#1 \kern0.1em{;}\kern0.1em #2\Bigr)}
\newcommand{\MIcond}[3]{\II\left(#1 \kern0.1em{;}\kern0.1em #2 \kern0.1em\middle|\kern0.1em #3\right)}
\newcommand{\eMIcond}[3]{\II(#1 \kern0.1em{;}\kern0.1em #2 \kern0.1em|\kern0.1em #3)} 
\newcommand{\bigMIcond}[3]{\II\bigl(#1 \kern0.1em{;}\kern0.1em #2 \kern-0.1em \bigm| \kern-0.1em#3\bigr)}
\newcommand{\BigMIcond}[3]{\II\Bigl(#1 \kern0.1em{;}\kern0.1em #2 \kern-0.1em \Bigm| \kern-0.1em#3\Bigr)}

\newcommand{\SMI}{\II_\infty}  
\newcommand{\SI}[2]{\SMI\left(#1 \kern0.1em{;}\kern0.1em #2\right)} 
\newcommand{\eSI}[2]{\SMI(#1 \kern0.1em{;}\kern0.1em #2)} 
\newcommand{\bigSI}[2]{\SMI\bigl(#1 \kern0.1em{;}\kern0.1em #2\bigr)}
\newcommand{\BigSI}[2]{\SMI\Bigl(#1 \kern0.1em{;}\kern0.1em #2\Bigr)}
\newcommand{\SIcond}[3]{\SMI\left(#1 \kern0.1em{;}\kern0.1em #2 \kern0.1em\middle|\kern0.1em #3\right)}
\newcommand{\eSIcond}[3]{\SMI(#1 \kern0.1em{;}\kern0.1em #2 \kern0.1em|\kern0.1em #3)} 
\newcommand{\bigSIcond}[3]{\SMI\bigl(#1 \kern0.1em{;}\kern0.1em #2 \kern-0.1em \bigm| \kern-0.1em#3\bigr)}
\newcommand{\BigSIcond}[3]{\SMI\Bigl(#1 \kern0.1em{;}\kern0.1em #2 \kern-0.1em \Bigm| \kern-0.1em#3\Bigr)}

\newcommand{\MaxL}{\mathop{}\!\mathsf{MaxL}}  
\newcommand{\ML}[2]{\MaxL\left(#1 \kern0.1em{;}\kern0.1em #2\right)} 
\newcommand{\eML}[2]{\MaxL(#1 \kern0.1em{;}\kern0.1em #2)} 
\newcommand{\bigML}[2]{\MaxL\bigl(#1 \kern0.1em{;}\kern0.1em #2\bigr)}
\newcommand{\BigML}[2]{\MaxL\Bigl(#1 \kern0.1em{;}\kern0.1em #2\Bigr)}
\newcommand{\MLcond}[3]{\MaxL\left(#1 \kern0.1em{;}\kern0.1em #2 \kern0.1em\middle|\kern0.1em #3\right)}
\newcommand{\eMLcond}[3]{\MaxL(#1 \kern0.1em{;}\kern0.1em #2 \kern0.1em|\kern0.1em #3)} 
\newcommand{\bigMLcond}[3]{\MaxL\bigl(#1 \kern0.1em{;}\kern0.1em #2 \kern-0.1em \bigm| \kern-0.1em#3\bigr)}
\newcommand{\BigMLcond}[3]{\MaxL\Bigl(#1 \kern0.1em{;}\kern0.1em #2 \kern-0.1em \Bigm| \kern-0.1em#3\Bigr)}

\newcommand{\relD}{\mathop{}\!\mathscr{D}}         
\newcommand{\relDf}[2]{\relD\left(#1 \kern0.1em\middle\|\kern0.1em #2\right)}
\newcommand{\erelDf}[2]{\relD(#1 \kern0.1em\|\kern0.1em #2)} 
\newcommand{\bigrelDf}[2]{\relD\bigl(#1 \kern-0.1em \bigm\| \kern-0.1em#2\bigr)}
\newcommand{\BigrelDf}[2]{\relD\Bigl(#1 \kern-0.1em \Bigm\| \kern-0.1em#2\Bigr)}
\newcommand{\biggrelDf}[2]{\relD\biggl(#1 \kern-0.1em \biggm\| \kern-0.1em#2\biggr)}
\newcommand{\BiggrelDf}[2]{\relD\Biggl(#1 \kern-0.1em \Biggm\| \kern-0.1em#2\Biggr)}

\newcommand{\Prscond}[2]{\Pr\left(#1 \kern0.1em\middle|\kern0.1em #2\right)}
\newcommand{\ePrscond}[2]{\Pr(#1 \kern0.1em|\kern0.1em #2)} 
\newcommand{\bigPrscond}[2]{\Pr\bigl(#1 \kern-0.1em \bigm| \kern-0.1em#2\bigr)}
\newcommand{\BigPrscond}[2]{\Pr\Bigl(#1 \kern-0.1em \Bigm| \kern-0.1em#2\Bigr)}
\newcommand{\biggPrscond}[2]{\Pr\biggl(#1 \kern-0.1em \biggm| \kern-0.1em#2\biggr)}
\newcommand{\BiggPrscond}[2]{\Pr\Biggl(#1 \kern-0.1em \Biggm| \kern-0.1em#2\Biggr)}

\newcommand{\ePrv}[1]{\Pr[#1]}

\newcommand{\Prvcond}[2]{\Pr\left[#1 \kern0.1em\middle|\kern0.1em #2\right]}
\newcommand{\ePrvcond}[2]{\Pr[#1 \kern0.1em|\kern0.1em #2]} 
\newcommand{\bigPrvcond}[2]{\Pr\bigl[#1 \kern-0.1em \bigm| \kern-0.1em#2\bigr]}
\newcommand{\BigPrvcond}[2]{\Pr\Bigl[#1 \kern-0.1em \Bigm| \kern-0.1em#2\Bigr]}
\newcommand{\biggPrvcond}[2]{\Pr\biggl[#1 \kern-0.1em \biggm| \kern-0.1em#2\biggr]}
\newcommand{\BiggPrvcond}[2]{\Pr\Biggl[#1 \kern-0.1em \Biggm| \kern-0.1em#2\Biggr]}

\newcommand{\Exp}{\operatorname{\textnormal{\textsf{E}}}}

\newcommand{\E}[2][]{\Exp_{#1}\left[#2\right]}

\newcommand{\bigE}[2][]{\Exp_{#1}\bigl[#2\bigr]}

\newcommand{\Econd}[3][]{\Exp_{#1}\left[#2 \kern0.1em\middle|\kern0.1em #3\right]}
\newcommand{\eEcond}[3][]{\Exp_{#1}[#2 \kern0.1em|\kern0.1em #3]}
\newcommand{\bigEcond}[3][]{\Exp_{#1}\bigl[#2 \kern-0.1em \bigm| \kern-0.1em #3\bigr]}
\newcommand{\BigEcond}[3][]{\Exp_{#1}\Bigl[#2 \kern-0.1em \Bigm| \kern-0.1em #3\Bigr]}
\newcommand{\biggEcond}[3][]{\Exp_{#1}\biggl[#2 \kern-0.1em \biggm| \kern-0.1em #3\biggr]}
\newcommand{\BiggEcond}[3][]{\Exp_{#1}\Biggl[#2 \kern-0.1em \Biggm| \kern-0.1em #3\Biggr]}

\newcommand{\Covcond}[4][]{\mathop{}\!\mathsf{Cov}_{#1}\left[{#2},{#3}\kern0.1em\middle|\kern0.1em{#4}\right]}
\newcommand{\eCovcond}[4][]{\mathop{}\!\mathsf{Cov}_{#1}[{#2},{#3}\kern0.1em|\kern0.1em{#4}]}
\newcommand{\bigCovcond}[4][]{\mathop{}\!\mathsf{Cov}_{#1}\bigl[{#2},{#3}\kern-0.1em\bigm|\kern-0.1em{#4}\bigr]}
\newcommand{\BigCovcond}[4][]{\mathop{}\!\mathsf{Cov}_{#1}\Bigl[{#2},{#3}\kern-0.1em\Bigm|\kern-0.1em{#4}\Bigr]}
\newcommand{\biggCovcond}[4][]{\mathop{}\!\mathsf{Cov}_{#1}\biggl[{#2},{#3}\kern-0.1em\biggm|\kern-0.1em{#4}\biggr]}
\newcommand{\BiggCovcond}[4][]{\mathop{}\!\mathsf{Cov}_{#1}\Biggl[{#2},{#3}\kern-0.1em\Biggm|\kern-0.1em{#4}\Biggr]}

\newcommand{\indep}{\mathrel{\bot}\joinrel\mathrel{\mkern-5mu}\joinrel\mathrel{\bot}}  
\newcommand{\markov}{\mathrel{\multimap}\joinrel\mathrel{-}\joinrel\mathrel{\mkern-6mu}\joinrel\mathrel{-}} 

\newcommand{\eqdef}{\triangleq} 

\makeatletter

\newcommand{\Rmnum}[1]{\expandafter\@slowromancap\romannumeral #1@}
\makeatother

\ifCLASSOPTIONcompsoc
    \usepackage[caption=false, font=normalsize, labelfont=sf, textfont=sf,subrefformat=parens,labelformat=parens]{subfig}
\else
\usepackage[caption=false, font=footnotesize,subrefformat=parens,labelformat=parens]{subfig}
\fi
\usepackage{lipsum} 
\usepackage[normalem]{ulem} 
\usepackage{balance}
\usepackage{amsbsy}
\usepackage{booktabs} 
\usepackage{enumitem}
\usepackage{mathtools}
\usepackage{amsmath,amssymb,amsfonts,amsthm} 
\usepackage{nicefrac}   
\usepackage[lined,boxed,commentsnumbered,linesnumbered,ruled]{algorithm2e}
\usepackage{comment}
\usepackage{pgfplots}
\pgfplotsset{compat=newest}
\usepackage{tikz} 
\usetikzlibrary{shapes,decorations.text,decorations.pathreplacing,arrows.meta,positioning,shadows,calc,patterns}

\newtheorem{theorem}{Theorem}
\newtheorem{lemma}{Lemma}
\newtheorem{corollary}{Corollary}

\newtheorem{proposition}{Proposition}

\newtheorem{definition}{Definition}
\newtheorem{remark}{Remark}

\usepackage[capitalize]{cleveref}
\crefname{equation}{\unskip}{\unskip} 
\crefname{claim}{Claim}{Claims} 
\usepackage{array}
\usepackage{multirow}
\newcolumntype{C}[1]{>{\centering\arraybackslash}p{#1}}
\allowdisplaybreaks

\renewcommand{\vect}[1]{\vectg{#1}} 
\newcommand{\collect}[1]{\mathscr{#1}} 
\newcommand{\Hwt}[1]{w_\mathsf{H}\left(#1\right)} 
\newcommand{\Spt}[1]{\chi\left(#1\right)} 
\newcommand*{\Resize}[2][4]{\resizebox{#1}{!}{\ensuremath{#2}}} 
\makeatletter
\renewcommand*\env@matrix[1][*\c@MaxMatrixCols c]{%
  \hskip -\arraycolsep
  \let\@ifnextchar\new@ifnextchar
  \array{#1}}
\makeatother

\renewcommand{\HH}{\mathop{}\!\mathsf{H}} 
\renewcommand{\II}{\mathop{}\!\mathsf{I}}  

\renewcommand{\Exp}{\operatorname{\mathbb{E}}}
\definecolor{darkgreen}{rgb}{0, 0.5, 0}

\begin{document}
%
\title{The Capacity of Single-Server Weakly-Private Information Retrieval}

\author{Hsuan-Yin~Lin,~\IEEEmembership{Senior Member,~IEEE}, Siddhartha~Kumar,~\IEEEmembership{Member,~IEEE}, Eirik~Rosnes,~\IEEEmembership{Senior~Member,~IEEE},~Alexandre~Graell~i~Amat,~\IEEEmembership{Senior~Member,~IEEE}, and Eitan~Yaakobi,~\IEEEmembership{Senior~Member,~IEEE}
  \thanks{This work was partially funded by the Swedish Research Council (grant \#2016-04253), the Israel Science Foundation (grant \#1817/18), and by the Technion Hiroshi Fujiwara Cyber Security Research Center and the Israel National Cyber Directorate. This article was presented in part at the IEEE International Symposium on Information Theory (ISIT), Los Angles, CA, USA, June 2020.}
\thanks{H.-Y.~Lin, S.~Kumar, and E.~Rosnes are with Simula UiB, N--5006 Bergen, Norway (e-mail: lin@simula.no; kumarsi@simula.no; eirikrosnes@simula.no).}
\thanks{A.~Graell i Amat is with the Department of Electrical Engineering, Chalmers University of Technology, SE--41296 Gothenburg, Sweden, and Simula UiB (e-mail: alexandre.graell@chalmers.se).}
\thanks{E.~Yaakobi is with the Department of Computer Science, Technion --- Israel Institute of Technology, Haifa, 3200003 Israel (email: yaakobi@gmail.com).} 
}


\maketitle

\begin{abstract}

A private information retrieval (PIR) protocol guarantees that a user can privately retrieve files stored in a database without revealing any information about the identity of the requested file. Existing information-theoretic PIR protocols ensure perfect privacy, i.e., zero information leakage to the servers storing the database, but at the cost of high download. In this work, we present weakly-private information retrieval (WPIR) schemes that trade off perfect privacy to improve the download cost when the database is stored on a single server. We study the tradeoff between the download cost and information leakage in terms of mutual information (MI) and maximal leakage (MaxL) privacy metrics. By relating the WPIR problem to rate-distortion theory, the download-leakage function, which is defined as the minimum required download cost of all single-server WPIR schemes for a given level of information leakage and a fixed file size, is introduced. By characterizing the download-leakage function for the MI and MaxL metrics, the capacity of single-server WPIR is fully described.

\end{abstract}


\section{Introduction}
\label{sec:introduction}


User privacy is becoming increasingly important both socially and politically in today's modern age of information (as demonstrated, for instance, by the European Union's General Data Protection Regulation).
In this context, private information retrieval (PIR), introduced by Chor \emph{et al.} in~\cite{ChorGoldreichKushilevitzSudan95_1}, has gained traction in the information theory community. In PIR, a user can retrieve a file from a database without revealing the identity of the file to the servers storing it. From an information-theoretic perspective, the file size is typically much larger than the size of the queries to all servers. Therefore rather than accounting for both the upload and the download cost, as is usually done in the computer science community, here efficiency is measured in terms of the download cost. More precisely, efficiency is measured in terms of PIR rate, which is the ratio between the requested file size and the total number of symbols downloaded. The supremum of the PIR rate over all possible schemes and over all file sizes 
is called the PIR capacity.

PIR was first addressed in the information theory literature by Shah \emph{et al.} \cite{ShahRashmiRamchandran14_1}, while the tradeoff between storage overhead and PIR rate was first considered in \cite{ChanHoYamamoto15_1}. 
Shortly after, Sun and Jafar~\cite{SunJafar17_1} characterized the PIR capacity for the classical PIR model of replicated servers. 
Since then the concept of PIR has been extended to several relevant scenarios: maximum distance separable (MDS) coded servers~\cite{TajeddineGnilkeElRouayheb18_1, BanawanUlukus18_1}, arbitrary linear coded servers~\cite{KumarLinRosnesGraellAmat19_1, LinKumarRosnesGraellAmat18_2, FreijGnilkeHollantiHTrautmannKarpukKubjas19_1}, colluding servers~\cite{SunJafar18_2, FreijGnilkeHollantiKarpuk17_1, TajeddineGnilkeElRouayheb18_1, KumarLinRosnesGraellAmat19_1, FreijGnilkeHollantiHTrautmannKarpukKubjas19_1, DOliveiraElRouayheb20_1, HolzbaurFreijLiHollanti19_1sub}, robust PIR~\cite{SunJafar18_2}, PIR with Byzantine servers~\cite{BanawanUlukus19_1}, optimal upload cost of PIR, i.e., the smallest query information required to be sent to the servers~\cite{TianSunChen19_1}, access complexity of PIR, i.e., the total number of symbols needed to be accessed across all servers in order to privately retrieve an arbitrary file~\cite{ZhangYaakobiEtzionSchwartz19_1}, single-server PIR with side information~\cite{KadheGarciaHeidarzadehElRouayhebSprintson20_1,HeidarzadehKazemiSprintson21_1}, PIR on graph-based replication systems~\cite{RavivTamoYaakobi20_1}, PIR with secure storage~\cite{YangShinLee18_1,JiaSunJafar19_1}, functional PIR codes~\cite{ZhangEtzionYaakobi20_1}, and private proximity retrieval codes~\cite{ZhangYaakobiEtzion19_2sub}.

Weakly-private information retrieval (WPIR)~\cite{LinKumarRosnesGraellAmatYaakobi19_1, SamyTandonLazos19_1,LinKumarRosnesGraellAmatYaakobi20_2sub} is an interesting extension of the original PIR problem as it allows for improvements in the download cost at the expense of some information leakage to the servers on the identity of the requested file.\footnote{On a related note, relaxing the requirement of perfect secrecy has been considered in the information theory literature in different contexts. For instance, in network coding, the term~\emph{weakly secure} is used when perfect security is only guaranteed for a subset of the messages multicasted from a source node~\cite{BhattadNarayanan05_1,SilvaKschischang09_1}. On the other hand,  \emph{weak security} in the context of secure communications refers to asymptotic \emph{per-symbol} zero information leakage~\cite{BlochBarros11_1}.} In particular,~\cite{LinKumarRosnesGraellAmatYaakobi19_1} considers the multi-server case with mutual information (MI) and worst-case information leakage~\cite{KopfBasin07_1} as privacy metrics, while  \cite{LinKumarRosnesGraellAmatYaakobi20_2sub} includes results for the maximum leakage (MaxL) privacy metric as well as converse bounds. In~\cite{SamyTandonLazos19_1}, Samy \emph{et al.}, under the name of leaky PIR, consider a privacy metric related to \emph{differential privacy}~\cite{DworkMcSherryNissimSmith06_1,Dwork06_1} for the multi-server case. The multi-server case under the MaxL  privacy metric has also been recently studied by Zhou \emph{et al.} \cite{ZhouGuoTian20_1}. 

In the computer science literature, to the best of our knowledge, there are only a few works that have considered relaxing the perfect privacy requirement of PIR in order to improve performance. In the first work \cite{AsonovFreytag02_1}, which appeared almost two decades ago, the perfect privacy condition was relaxed by introducing the concept of \emph{repudiation}.  A protocol assures the repudiation property if the probability of all designed queries to retrieve any file stored in the database is strictly smaller than one. Hence, the user can deny any claim about the identity of the desired file by the server. However, the condition of repudiation can be achieved even if the server can determine the identity of the requested file almost surely, and thus, it does not provide a good level of information-theoretic privacy. More recently, Toledo \emph{et al.} \cite{ToledoDanezisGoldberg16_1} adopted a privacy metric based on differential privacy in order to enhance the efficiency of PIR by lowering the level of privacy. Moreover, both  works did not study any fundamental information-theoretic tradeoffs between information leakage and different costs under the considered privacy metrics. In contrast to WPIR, where the information leakage to the servers is considered, recently, the authors of~\cite{GuoZhouTian20_1} studied the information leakage of the nondesired files to the user in PIR systems. Furthermore, along the same lines of research, leaky PIR was extended to the symmetric PIR setting in~\cite{SamyAttiaTandonLazos20_2sub}. Symmetric PIR is a variant of PIR where in addition to the privacy request, the user cannot learn anything about the remaining files in the database when the user retrieves its desired file~\cite{SunJafar19_1}. In~\cite{SamyAttiaTandonLazos20_2sub}, the symmetric PIR requirement of zero information leakage on the nondesired files as well as the perfect privacy requirement on the identity of the requested file are both relaxed in order to improve the download rate.

In distributed storage systems (DSSs), for several applications, it is more realistic to assume that all servers can collude. In the single server scenario, it can be shown that all files stored in the database need to be downloaded to guarantee perfect privacy~\cite{ChorKushilevitzGoldreichSudan98_1}. This implies that the PIR rate tends to zero as the number of stored files increases. In \cite{KadheGarciaHeidarzadehElRouayhebSprintson20_1}, the authors introduce PIR schemes that improve the download cost by leveraging on the assumption that the user has some prior side information on the content of the database. Two cases are considered, namely whether or not the privacy of the side information needs to be preserved. Lastly, latent-variable PIR for the single server setting was introduced in~\cite{SamyAttiaTandonLazos20_1}. The goal of latent-variable PIR is to completely hide the latent attributes induced by the requested file identity. The latent-variable PIR framework can be seen as a variant of WPIR as no leakage on the latent attributes can be achieved even if parts of the identity of the requested file are leaked. 

In this paper, we relax the condition of perfect privacy in the single server setting.
In similar lines to~\cite{LinKumarRosnesGraellAmatYaakobi19_1}, we show that by relaxing the perfect privacy requirement, the download cost can be improved. Like~\cite{LinKumarRosnesGraellAmatYaakobi20_2sub}, we consider both the MI and MaxL privacy metrics~\cite{Smith09_1,BartheKopf11_1,IssaWagnerKamath20_1}, where the latter is the most robust information-theoretic metric for information leakage yet known. In particular, we establish a connection between the single-server WPIR problem and rate-distortion theory, which provides fundamental insights to describe the optimal tradeoff between the download cost and the allowed information leakage. The primary contribution of this work is to characterize the capacity (defined as the inverse of the minimum download cost over all possible schemes and over all file sizes) of single-server WPIR when the information leakage to the server is measured in terms of MI or MaxL. In this work, the minimum achievable download cost for a given information leakage constraint and for an \emph{arbitrary} fixed file size is determined, and thus the WPIR capacity 
is derived. Especially, we propose a simple novel single-server WPIR scheme that achieves the WPIR capacity for both the MI and MaxL privacy metrics. 
Finally, we remark here that also the notion of differential privacy can be adapted to our setting, e.g., the local differential privacy metric~\cite{Kasiviswanathan-etal08_1,DuchiJordanWainwright13_1}. However, since the local differential privacy metric normally provides \emph{stronger}  privacy guarantees than the MI or MaxL metrics, it can readily be shown that it is not possible to further lower the download cost in the single server scenario. This is in contrast to the case of multiple servers~\cite{SamyTandonLazos19_1}.

The remainder of this paper is structured as follows. Section~\ref{sec:preliminaries-ProblemStatement} presents the notation, definitions, and the problem formulation. In Section~\ref{sec:characterizaiton_download-leakage}, we introduce the download-leakage function of single-server WPIR, which is defined as the minimum achievable download cost for a given information leakage constraint and for an arbitrary file size. Moreover, we discuss some properties of the function when the leakage is measured in terms of the MI or MaxL metrics. In Section~\ref{sec:WPIR-schemes_basic-partition}, a basic solution for single-server WPIR is presented in which the file indices are partitioned into several partitions. In Section~\ref{sec:capacity_single-server-WPIR}, we give a closed-form expression for the single-server WPIR capacity for both the MI and MaxL metrics. A capacity-achieving WPIR scheme is proposed in Section~\ref{sec:achievability}. The converse result on the minimum download cost for the MI metric is provided in Section~\ref{sec:converse_Thm1}, while that of the MaxL metric is given in Section~\ref{sec:converse_Thm2}. Finally, Section~\ref{sec:conclusion} concludes the paper.

\section{Preliminaries and Problem Statement}
\label{sec:preliminaries-ProblemStatement}

\subsection{Notation}
\label{sec:notation-definitions}

We denote by $\Naturals$ the set of all positive integers, $[a]\eqdef\{1,2,\ldots,a\}$, and $[a:b]\eqdef\{a,a+1,\ldots,b\}$ for $a,b\in\{0\}\cup\Naturals$ and $a \leq b$. The set of nonnegative real numbers is denoted by $\Reals_{+}$. Vectors are denoted by bold letters, matrices by sans serif capital letters, and sets by calligraphic uppercase letters, e.g., $\vect{x}$, $\mat{X}$, and $\set{X}$, respectively. In general, vectors are represented as row vectors throughout the paper. We use uppercase letters for random variables (RVs) (either scalar or vector), e.g., $X$ or $\vect{X}$.
For a given index set $\set{S}$, we write $X^\set{S}$ to represent $\bigl\{X^{(m)}\colon m\in\set{S}\bigr\}$. $X\indep Y$ means that the two RVs $X$ and $Y$ are independent. $\trans{(\cdot)}$ denotes the transpose of its argument. The Hamming weight of a vector $\vect{x}$ is denoted by $\Hwt{\vect{x}}$, while its support will be denoted by $\Spt{\vect{x}}$. $\E[X]{\cdot}$ and $\E[P_X]{\cdot}$ denote expectation with respect to the RV $X$ and distribution $P_X$, respectively. $\HP{X}$, $\HP{P_X}$, or $\HP{p_{1},\ldots,p_{\card{\set{X}}}}$ represents the entropy of $X$, where $P_{X}(\cdot)=(p_1,\ldots,p_{\card{\set{X}}})$ denotes the distribution of the RV $X$. $\eMI{X}{Y}$ denotes the MI between $X$ and $Y$. 

\subsection{System Model}
\label{sec:system-model}

We consider a single server that stores $\const{M}$ independent files $\vect{X}^{(1)},\ldots,\vect{X}^{(\const{M})}$, where each file $\vect{X}^{(m)}=\bigl(X_1^{(m)},\ldots,X_\beta^{(m)}\bigr)$, $m\in [\const{M}]$, is represented as a length-$\beta$ row vector over $\set{X}$. Assume that each element of $\vect{X}^{(m)}$ is chosen independently and uniformly at random from $\set{X}$. Thus, we have $\bigHP{\vect{X}^{(m)}}=\beta\log_2{\card{\set{X}}}$ bits, $\forall\,m\in [\const{M}]$. A user wishes to efficiently retrieve $\vect{X}^{(M)}$ by allowing some information leakage to the server, where the requested file index $M$ is assumed to be uniformly distributed over $[\const{M}]$.\footnote{Note that the requested file index $M$ does not necessarily need to be uniformly distributed, which is referred to as semantic PIR in the literature \cite{VithanaBanawanUlukus20_1sub}.} 
Similar to the detailed mathematical description in~\cite{LinKumarRosnesGraellAmatYaakobi19_1}, we give the following definition for a single-server WPIR scheme.
\begin{definition}  
  \label{def:WPIRscheme_single-server}
  An $\const{M}$-file WPIR scheme $\collect{C}$ for a single server storing $\const{M}$ files consists of:
  \begin{itemize}
  \item A random strategy $\vect{S}$ with alphabet $\set{S}$, which is privately designed by the user.
    
  \item A query function 
    \begin{IEEEeqnarray*}{rCl}
      \phi\colon\{1,\ldots,\const{M}\}\times\set{S}\to\set{Q}
    \end{IEEEeqnarray*}
    that generates a query $\vect{Q}=\phi(M,\vect{S})$ with alphabet $\set{Q}$, and induces a conditional probability mass function (PMF) $P_{\vect{Q}|M}$. The query $\vect{Q}$ is sent to the server to retrieve the $M$-th file.
  \item An answer function
    \begin{IEEEeqnarray*}{rCl}
      \varphi\colon\set{Q}\times\set{X}^{\beta\const{M}}\to\set{A}^{\beta L}
    \end{IEEEeqnarray*}
    that returns the answer $\vect{A}=\varphi(\vect{Q},\vect{X}^{[\const{M}]})$ back to the user, with download symbol alphabet $\set{A}$. Here, $L=L(\vect{Q})$ is the normalized length of the answer, which is a function of the query $\vect{Q}$.\footnote{Note that in this work, the performance metric we focus on is the normalized download cost (see~\eqref{eq:D-def_single-serverWPIR} later). Hence, without loss of generality, we define the answer-length in a normalized manner.} More specifically, given a query realization $\vect{Q}=\vect{q}$, $L(\vect{q})$ can be seen as the  codeword length of a code that encodes the files $\vect{X}^{[\const{M}]}$, which is independent of the particular realization of the files.\footnote{
      From a source coding perspective, the files/sources are encoded by a \emph{fixed-length} code, i.e., $L(\vect{q})$ is independent of the realization of the files, reflecting the fact that the files are independent and identically distributed (i.i.d.) according to a uniform distribution. As opposed to a \emph{variable-length} code, for a fixed-length code all codewords are of equal length. Note that this setup follows the problem formulation in the PIR literature, see, e.g.,~\cite{TianSunChen19_1}.}
  \item A privacy leakage metric $\rho^{(\cdot)}(P_{\vect{Q}|M})\geq 0$, which is defined as a function of $P_{\vect{Q}|M}$, that measures the amount of leaked information of the identity of the requested file to the server by observing the generated query $\vect{Q}$, where the superscript ``$(\cdot)$'' indicates the used metric.
  \end{itemize}
  Furthermore, the scheme should allow a user to retrieve the requested file from the answer, the query, the index of the requested file, and the random strategy. In other words, this scheme must satisfy the condition of perfect retrievability,
  \begin{IEEEeqnarray}{rCl}
    \bigHPcond{\vect{X}^{(M)}}{\vect{A},\vect{Q},M,\vect{S}}=0.
    \label{eq:retrievability}
  \end{IEEEeqnarray}
\end{definition}
We remark that a PIR scheme corresponds to a WPIR scheme for which no information leakage is allowed.

\subsection{Metrics of Information Leakage}
\label{sec:metrics_information-leakage}

Given a single-server $\const{M}$-file WPIR scheme and a fixed distribution $P_M$, the conditional PMF of the query given the index $M$ of the requested file, $P_{\vect{Q}|M}$, can be seen as a privacy mechanism (a randomized mapping). The server receives the random outcome $\vect{Q}$ of the privacy mechanism $P_{\vect{Q}|M}$, and is curious about the index $M$ of the requested file. The information leakage of a WPIR scheme is then measured with respect to its corresponding privacy mechanism $P_{\vect{Q}|M}$.

In this paper, we focus on two commonly-used information-theoretic measures, namely MI and MaxL. For the former, the information leakage is quantified by
\begin{IEEEeqnarray}{c}
  \rho^{(\mathsf{MI})}(P_{\vect{Q}|M})\eqdef\eMI{M}{\vect{Q}}.
  \label{eq:expression_MI}
\end{IEEEeqnarray}
The second privacy metric, MaxL, which is introduced in~\cite{BartheKopf11_1,IssaWagnerKamath20_1}, is quantified by 
\begin{IEEEeqnarray}{rCl}
  \rho^{(\mathsf{MaxL})}(P_{\vect{Q}|M})& \eqdef &\ML{M}{\vect{Q}}
  \nonumber\\
  & = &\log_2{\sum_{\vect{q}\in\set{Q}}\max_{m\in[\const{M}]}}P_{\vect{Q}|M}(\vect{q}|m).\label{eq:expression_MaxL}
\end{IEEEeqnarray}
We remark that MaxL can also be defined based on the \emph{min-entropy (MinE) information leakage} $\SI{M}{\vect{Q}}$ for the privacy mechanism $P_{\vect{Q}|M}$, where
\begin{IEEEeqnarray*}{c}
  \SI{M}{\vect{Q}}\eqdef\MH{M}-\MH{M|\vect{Q}}
\end{IEEEeqnarray*}
and $\eMH{M}$ denotes the MinE measure that is widely discussed in the computer science literature, see, e.g.,~\cite{Smith09_1}. On the other hand, the authors in~\cite{IssaWagnerKamath20_1} proposed the equivalent definition 
\begin{IEEEeqnarray*}{c}
  \ML{M}{\vect{Q}}\eqdef\sup_{X\markov M\markov \vect{Q}\markov\hat{X}}\log_2{\frac{\ePrv{X=\hat{X}}}{\max_{x\in\set{X}}P_X(x)}}
\end{IEEEeqnarray*}
of MaxL, where the supremum is taken over all possible $X$ and $\hat{X}$ taking values in the same finite, but arbitrary alphabet $\set{X}$, and the notation $X\markov M\markov \vect{Q}\markov\hat{X}$ means that the RVs $X$, $M$, $\vect{Q}$, and $\hat{X}$ form a Markov chain from left to right. Note that if $\ML{M}{\vect{Q}}=\rho$ bits, the above definition indicates that for any possible randomized function $X$ of $M$, the maximum probability of correctly guessing $X$ based on $\vect{Q}$ is bounded from above by the product of $2^{\rho}$ and the maximum probability of guessing $X$ with no observation. 


It is worth mentioning that since we assume that $M$ is uniformly distributed, the MinE information leakage and the MaxL privacy metric can be shown to be equivalent, i.e., $\ML{M}{\vect{Q}}=\SI{M}{\vect{Q}}$~\cite[Th.~1]{BartheKopf11_1}, \cite[Th.~1]{IssaWagnerKamath20_1}. 
It is also worth mentioning that there is a relation between MaxL and differential privacy, see, e.g.,~\cite[Th.~3]{BartheKopf11_1}.

The following lemma summarizes some useful properties for both the MI and MaxL privacy metrics.
\begin{lemma}[Data processing inequalities {\cite[Lem.~1, Cor.~1]{IssaWagnerKamath20_1}}]
  \label{lem:convex_MI-MaxL}
  For any joint distribution $P_{X,Y}$,
  \begin{enumerate}
  \item if the RVs $X,Y$, and $Z$ form a Markov chain, then
    \begin{IEEEeqnarray*}{rCl}
      \MI{X}{Z}& \leq &\min\{\MI{X}{Y},\MI{Y}{Z}\},\textnormal{ and}
      \\
      \ML{X}{Z}& \leq &\min\{\ML{X}{Y},\ML{Y}{Z}\}.
    \end{IEEEeqnarray*}
  \item Consider a fixed distribution $P_X$. Then, both $\MI{X}{Y}$ and $2^{\ML{X}{Y}}$ are convex functions in $P_{Y|X}$.
  \end{enumerate}
\end{lemma}

Throughout the paper, the information leakage metric of a WPIR scheme $\collect{C}$ is denoted by $\rho^{(\cdot)}(\collect{C})$. Moreover, since $\rho^{(\cdot)}$ is defined with a single argument of $P_{\vect{Q}|M}$, we will also simply write the corresponding MI and MaxL metrics as a function of $P_{\vect{Q}|M}$, i.e., $\II(P_{\vect{Q}|M})=\MI{M}{\vect{Q}}$ and $\MaxL(P_{\vect{Q}|M})=\ML{M}{\vect{Q}}$.

\subsection{Download Cost and Rate for a Single-Server WPIR Scheme}
\label{sec:achievable-rate_IR}

From the perfect privacy requirement of PIR, the server should not be able to differentiate the returned answers (e.g., by looking at their sizes) no matter which file index is requested. However, in contrast to PIR, the download cost for WPIR may be different for the retrieval of different files. Hence, the download cost is defined as the expected download cost over all possible requested files. The download cost of a single-server WPIR scheme $\collect{C}$ for the retrieval of the $m$-th file, denoted by $\const{D}^{(m)}(\collect{C})$, is defined as the normalized expected length of the returned answer over all random queries,
\begin{IEEEeqnarray*}{c}
  \const{D}^{(m)}(\collect{C})\eqdef\frac{\log_2{\card{\set{A}}}\E[P_{\vect{Q}|M=m}]{L(\vect{Q})}}{\log_2{\card{\set{X}}}},
\end{IEEEeqnarray*}
and the overall download cost, denoted by $\const{D}(\collect{C})$, is measured in terms of the normalized expected download cost over all files, i.e.,
\begin{IEEEeqnarray}{rCl}
  \const{D}(\collect{C})& \eqdef &\frac{\log_2{\card{\set{A}}}\E[P_{M}]{\E[P_{\vect{Q}|M}]{L(\vect{Q})}}}{\log_2{\card{\set{X}}}}
  \nonumber\\
  & = &\gamma\E[M,\vect{Q}]{L(\vect{Q})},
  \IEEEeqnarraynumspace\label{eq:D-def_single-serverWPIR}
\end{IEEEeqnarray}%
where $\gamma\eqdef\nicefrac{\log_2{\card{\set{A}}}}{\log_2{\card{\set{X}}}}$. We remark that in general the sizes of the download symbol and file symbol alphabets can be different. However, for simplicity, we assume $\gamma=1$ throughout this paper as the value of $\gamma$ does not affect the generality of the results. 
Accordingly, the WPIR rate is defined as $\const{R}(\collect{C})\eqdef\inv{\const{D}(\collect{C})}$.

Intuitively, a smaller download cost can be achieved if we allow a higher level of information leakage. In this paper, our goal is to characterize the optimal tradeoff between the download cost and the allowed information leakage with respect to a privacy metric. We start with the following definition of an achievable download-leakage pair.
\begin{definition}
  \label{def:download-leakage-pair}
  Consider a single server that stores $\const{M}$ files. A download-leakage pair $(\const{D},\varrho)$ is said to be \emph{achievable} in terms of the information leakage metric $\rho^{(\cdot)}$ if there exists a WPIR scheme $\collect{C}$ such that $\E[M,\vect{Q}]{L(\vect{Q})}\leq\const{D}$ and $\rho^{(\cdot)}(\collect{C})\leq\varrho$. The \emph{download-leakage} region is the set of all achievable download-leakage pairs $(\const{D},\varrho)$.
\end{definition}

\begin{remark}
  \label{remark:achievable_D-rho}
  By Definition~\ref{def:download-leakage-pair}, it is clear that if the pair $(\const{D},\varrho)$ is achievable, then the pair $(\const{D}',\varrho')$ with $\const{D}'\geq\const{D}$ and $\varrho'\geq\varrho$ is also achievable.
\end{remark}

\section{Characterization of the Optimal Download-Leakage Tradeoff}
\label{sec:characterizaiton_download-leakage}

Consider the single-server WPIR problem with an arbitrary file size $\beta$, where the leakage is measured by $\rho^{(\mathsf{MI})}$ or $\rho^{(\mathsf{MaxL})}$. The minimum achievable download cost for a given leakage constraint $\varrho$ can be formulated as the optimization problem 
\begin{IEEEeqnarray}{rCl}
  \IEEEyesnumber\label{eq:download-leakage-ft}
  \IEEEyessubnumber*
  \textnormal{minimize} & &\qquad\E[P_{M}P_{\vect{Q}|M}]{L(\vect{Q})}
  \\
  \textnormal{subject to} && \qquad \{L(\vect{q})\}_{\vect{q}\in\set{Q}}\subseteq\set{L}_\mathsf{ret},
  \label{eq:answers_retrievability}\\[1mm]
  & &\qquad\rho^{(\cdot)}(P_{\vect{Q}|M})\leq\varrho \label{eq:PMFs_leakage-constraint}
\end{IEEEeqnarray}
over the query function $\phi(\cdot)$ and the answer function $\varphi(\cdot)$ from Definition~\ref{def:WPIRscheme_single-server}, 
where $\set{L}_\mathsf{ret}$ is defined as the set of the codeword lengths of all possible fixed-length codes that satisfy~\eqref{eq:retrievability}. In this way, it involves the query function, the answer generating function, and the decoding function from Definition~\ref{def:WPIRscheme_single-server}. For brevity, the explicit query and answer function constraints are omitted as $\set{L}_\mathsf{ret}$ implicitly involves these functions. We recall Definition~\ref{def:WPIRscheme_single-server} here that given a query realization $\vect{Q}=\vect{q}$, $L(\vect{q})$ can be seen as the codeword length of a fixed-length code that encodes the files and satisfies the {\itshape lossless} property in \eqref{eq:retrievability}. Note again that the fixed-length assumption for the code, or equivalently that $L(\vect{q})$ is independent of the specific realization of the files, reflects the fact that the files are i.i.d.~according to a uniform distribution. Finally, we remark that since $P_M$ is assumed to be fixed, the minimization over $P_M P_{\vect{Q}|M}$ in~\eqref{eq:download-leakage-ft} is taken over the set of all conditional distributions $P_{\vect{Q}|M}$.

\subsection{The Download-Leakage Function for Single-Server WPIR}
\label{sec:download-leakage-function}

To characterize the optimal achievable pairs of download cost and information leakage, we define two functions that describe the boundary of the download-leakage region.
\begin{definition}
  \label{def:download-leakage_SWPIR}
  For any file size $\beta$, the download-leakage function $\const{D}^{(\cdot)}(\varrho)$ for single-server WPIR is the minimum of all possible download costs $\const{D}$ for a given information leakage constraint $\varrho$ such that $(\const{D},\varrho)$ is achievable, i.e.,
  \begin{IEEEeqnarray*}{c}
    \const{D}^{(\cdot)}(\varrho)\eqdef\!\min_{\{L(\vect{q})\}_{\vect{q}\in\set{Q}}\subseteq\set{L}_\mathsf{ret},\,P_{\vect{Q}|M}\colon\rho^{(\cdot)}(P_{\vect{Q}|M})\leq\varrho}\!\E[P_{M}P_{\vect{Q}|M}]{L(\vect{Q})}.\IEEEeqnarraynumspace
  \end{IEEEeqnarray*}
\end{definition}
Following the notion of information-theoretic PIR capacity in the literature, we define the single-server WPIR capacity as the supremum of the inverse of all  download-leakage functions over all possible values of $\beta$ as follows.
\begin{definition}
    The single-server WPIR capacity is defined as $\const{C}^{(\cdot)}(\varrho)=\max_{\beta\in\Naturals}\bigl\{\inv{\bigl[\const{D}^{(\cdot)}(\varrho)\bigr]}\bigr\}$. 
\end{definition}
We remark here that the optimality results for the download-leakage functions in Sections~\ref{sec:converse_Thm1} and~\ref{sec:converse_Thm2} hold for any fixed file size $\beta$, since the solutions of the convex optimization problem formulations are independent of $\beta$, which indicates $\const{C}^{(\cdot)}(\varrho)=\inv{\bigl[\const{D}^{(\cdot)}(\varrho)\bigr]}$ for any file size $\beta$. This also implies that increasing the file size does not further improve the performance.
Note that it is known from the original work of Chor \emph{et al.}~\cite{ChorKushilevitzGoldreichSudan98_1} that the single-server PIR capacity is $\const{C}_\const{M} = \frac{1}{\const{M}}$.

We assume throughout the rest of the paper that the perfect retrievability condition in \eqref{eq:answers_retrievability} holds. Hence, for convenience, we will sometimes drop the condition of~\eqref{eq:answers_retrievability} in the download-leakage optimization formulation.

Naturally, we can also determine the optimal download-leakage region by interchanging the roles of the download cost and the information leakage.
\begin{definition}
  \label{def:leakage-download_SWPIR}
  For any file size $\beta$, the leakage-download function $\rho^{(\cdot)}(\const{D})$ for single-server WPIR is the minimum of all possible information leakages $\varrho$ for a given download cost constraint $\const{D}$ such that $(\const{D},\varrho)$ is achievable.  
\end{definition}  

\begin{lemma}
  \label{lem:convexity_download-leakage-ft}
  \begin{enumerate}
  \item For any file size $\beta$, the MI download-leakage function
  \begin{IEEEeqnarray*}{c}
    \const{D}^{(\mathsf{MI})}(\varrho)=\min_{P_{\vect{Q}|M}\colon\II(P_{\vect{Q}|M})\leq\varrho}\E[P_{M}P_{\vect{Q}|M}]{L(\vect{Q})}
  \end{IEEEeqnarray*}
  is convex in $\varrho$.
  \item For any file size $\beta$, the MaxL download-leakage function
  \begin{IEEEeqnarray*}{c}
    \const{D}^{(\mathsf{MaxL})}(\varrho)=\min_{P_{\vect{Q}|M}\colon\MaxL(P_{\vect{Q}|M})\leq\varrho}\E[P_{M}P_{\vect{Q}|M}]{L(\vect{Q})}
  \end{IEEEeqnarray*}
  is not a convex function, but $\const{D}^{(\mathsf{MaxL})}(\log_2(\varrho))$ is convex in $\varrho$.
  \end{enumerate}
\end{lemma}

\begin{IEEEproof}
  We first prove the lemma for MI leakage. Assume that $\const{D}^{(\mathsf{MI})}(\varrho_1)=\E[P_{M}P_{\vect{Q}_1^\ast|M}]{L(\vect{Q})}$ and $\const{D}^{(\mathsf{MI})}(\varrho_2)=\E[P_{M}P_{\vect{Q}_2^\ast|M}]{L(\vect{Q})}$ are achieved by the answer-lengths and conditional distributions $\bigl(\{L(\vect{q})\}_{\vect{q}\in\set{Q}^\ast_1},P_{\vect{Q}^\ast_1|M}\bigr)$ and $\bigl(\{L(\vect{q})\}_{\vect{q}\in\set{Q}^\ast_2},P_{\vect{Q}^\ast_2|M}\bigr)$, respectively, where $\MI{M}{\vect{Q}^\ast_1}\leq\varrho_1$ and $\MI{M}{\vect{Q}^\ast_2}\leq\varrho_2$. Let $P_{\vect{Q}_\lambda|M}$ be the distribution 
  \begin{IEEEeqnarray*}{c}
    P_{\vect{Q}_\lambda|M}=(1-\lambda)P_{\vect{Q}^{\ast}_1|M}+\lambda P_{\vect{Q}^{\ast}_2|M},
  \end{IEEEeqnarray*}
  defined over $\set{Q}_\lambda\eqdef\{(i,\vect{q}_i)\colon \vect{q}_i\in\set{Q}_i,i=1,2\}$. It can be seen that $\{L(\vect{q})\}_{\vect{q}\in\set{Q}_\lambda}\subseteq\set{L}_\mathsf{ret}$.

  Observe that since $\MI{M}{\vect{Q}}$ is convex in $P_{\vect{Q}|M}$, it follows that $\II\bigl((1-\lambda)P_{\vect{Q}^{\ast}_1|M}+\lambda P_{\vect{Q}^{\ast}_2|M}\bigr)\leq (1-\lambda)\II\bigl(P_{\vect{Q}^{\ast}_1|M}\bigr)+\lambda\II\bigl(P_{\vect{Q}^{\ast}_2|M}\bigr) \leq (1-\lambda)\varrho_1+\lambda\varrho_2$, which implies that $P_{\vect{Q}_\lambda|M}$ is an element of $\{P_{\vect{Q}|M}\colon\II(P_{\vect{Q}|M})\leq(1-\lambda)\varrho_1+\lambda\varrho_2\}$.
  
  Thus, by definition we get
  \begin{IEEEeqnarray}{rCl}
    \IEEEeqnarraymulticol{3}{l}{%
      \const{D}^{(\mathsf{MI})}((1-\lambda)\varrho_1+\lambda\varrho_2)}\nonumber\\*\quad%
    & = &\min_{P_{\vect{Q}|M}\colon\II(P_{\vect{Q}|M})\leq(1-\lambda)\varrho_1+\lambda\varrho_2}\E[P_{M}P_{\vect{Q}|M}]{L(\vect{Q})}
    \nonumber\\
    & \leq &\E[P_{M}P_{\vect{Q}_\lambda|M}]{L(\vect{Q})}
    \nonumber\\
    & \stackrel{(a)}{=} &(1-\lambda)\E[P_{M}P_{\vect{Q}_1^\ast|M}]{L(\vect{Q})}+\lambda\E[P_{M}P_{\vect{Q}_2^\ast|M}]{L(\vect{Q})}
    \label{eq:use_Plambda}\nonumber\\
    & = &(1-\lambda)\const{D}^{(\mathsf{MI})}(\varrho_1)+\lambda\const{D}^{(\mathsf{MI})}(\varrho_2),\nonumber 
  \end{IEEEeqnarray}
  where $(a)$ follows directly from the definition of $P_{\vect{Q}_\lambda|M}$. This shows that $\const{D}^{(\mathsf{MI})}(\varrho)$ is convex in $\varrho$ for the MI metric. The proof for the MaxL metric is analogous, since $2^{\MaxL(P_{\vect{Q}|M})}$ is convex in $P_{\vect{Q}|M}$.
\end{IEEEproof}
From Remark \ref{remark:achievable_D-rho} we can see that the convexity of the download-leakage function is very useful, since it can help to describe the download-leakage region if some achievable pairs are known. This observation can be summarized in the following corollary.
\begin{corollary}
  \label{cor:convexity_achievable-pairs}
  Assume that both pairs $(\const{D}_1,\varrho_1)$ and $(\const{D}_2,\varrho_2)$ are achievable. Then, for any $\lambda\in[0,1]$, the pair $\bigl(\const{D}_\lambda=(1-\lambda)\const{D}_1+\lambda\const{D}_2,\varrho_\lambda=(1-\lambda)\varrho_1+\lambda\varrho_2\bigr)$ is achievable under MI leakage, while the pair $\bigl(\const{D}_\lambda=(1-\lambda)\const{D}_1+\lambda\const{D}_2,\varrho_\lambda=\log_2{[(1-\lambda)2^{\varrho_1}+\lambda2^{\varrho_2}]}\bigr)$ is achievable for MaxL.
\end{corollary}

\subsection{Connection to Rate-Distortion Theory}
\label{sec:connection_rate-distortion-theory}

The celebrated \emph{rate-distortion theory} of Shannon and Kolmogorov (see, e.g.,~\cite[Ch.~9]{Gallager68_1},~\cite[Ch.~10]{moser19_v43}, and references therein) determines the minimum source compression rate required to reproduce any source sequence under a fidelity constraint, which is provided through a \emph{distortion measure} between the source sequence and the reconstructed sequence. Consider an information source sequence with i.i.d. 
 components according to $P_X$ and a distortion measure $d(\vect{x},\hat{\vect{x}})$ between the source sequence $\vect{x}$ and the reconstructed sequence $\hat{\vect{x}}$. The optimal rate-distortion region is characterized by the \emph{rate-distortion function}, defined as the minimum achievable compression rate $\eMI{X}{\hat{X}}$ under a given constraint on the average distortion $\bigE[P_{X}P_{\hat{X}|X}]{d(X,\hat{X})}$, where $\hat{X}$ represents the reconstructed source.


One important observation from~\eqref{eq:D-def_single-serverWPIR} is that, if we add the desired file index $m$ as an argument to the answer-length function $L$ by defining $L(m,\vect{q})\eqdef L(\vect{q})$ for all $m\in[\const{M}]$ for which $P_{\vect{Q}|M}(\vect{q}|m)>0$, and $L(m,\vect{q})\eqdef\infty$ otherwise (i.e., an infinite length for a given $m$ and query realization $\vect{q}$ indicates that $\vect{q}$ is never sent when requesting the $m$-th file), then the download cost can be expressed as $\E[P_{M}P_{\vect{Q}|M}]{L(\vect{Q})}=\E[P_{M}P_{\vect{Q}|M}]{L(M,\vect{Q})}$. 
Thus, in terms of the MI privacy metric, the leakage-download function of a given WPIR scheme can be related to the rate-distortion function, where the leakage and the download cost play similar roles as the compression rate and the average distortion, respectively. Below, we will equivalently use either $L(\vect{q})$ or $L(m,\vect{q})$ (as defined above). We defer the detailed discussion to Section~\ref{sec:converse_Thm1} where we will utilize results from rate-distortion theory to characterize the optimal leakage-download tradeoff for single-server WPIR.

\section{Results}
\label{sec:results}

\subsection{Partition WPIR Scheme}
\label{sec:WPIR-schemes_basic-partition}

In~\cite{LinKumarRosnesGraellAmatYaakobi19_1}, a WPIR scheme based on partitioning was proposed. The set of all file indices, i.e., $[\const{M}]$, is first privately pre-partitioned into $\eta$ equally-sized partitions by the user, each consisting of $\const{M}_{\eta}$ file indices, where $\const{M}_{\eta}=\const{M} / \eta \in\Naturals$. If there exists a viable $\const{M}_{\eta}$-file WPIR scheme, the user can apply the $\const{M}_{\eta}$-file WPIR scheme as a subscheme on each partition, and retrieve a file from the corresponding partition.

The partition $\const{M}$-file WPIR scheme is formally described as follows. Assume that the requested file $\vect{X}^{(m)}$ belongs to the $j$-th partition, where $j\in[\eta]$. Then, the query $\vect{Q}$ is constructed as
\begin{IEEEeqnarray}{c}
  \vect{Q}=\bigl(\widetilde{\vect{Q}},j\bigr)\in\tilde{\set{Q}}\times [\eta],
  \label{eq:queries_partition-scheme}
\end{IEEEeqnarray}
where $\widetilde{\vect{Q}}$ is the query of an existing $\const{M}_{\eta}$-file WPIR scheme. 

\begin{figure*}[t!]
  \subfloat[]{\includegraphics[width=0.5\textwidth]{\Figs/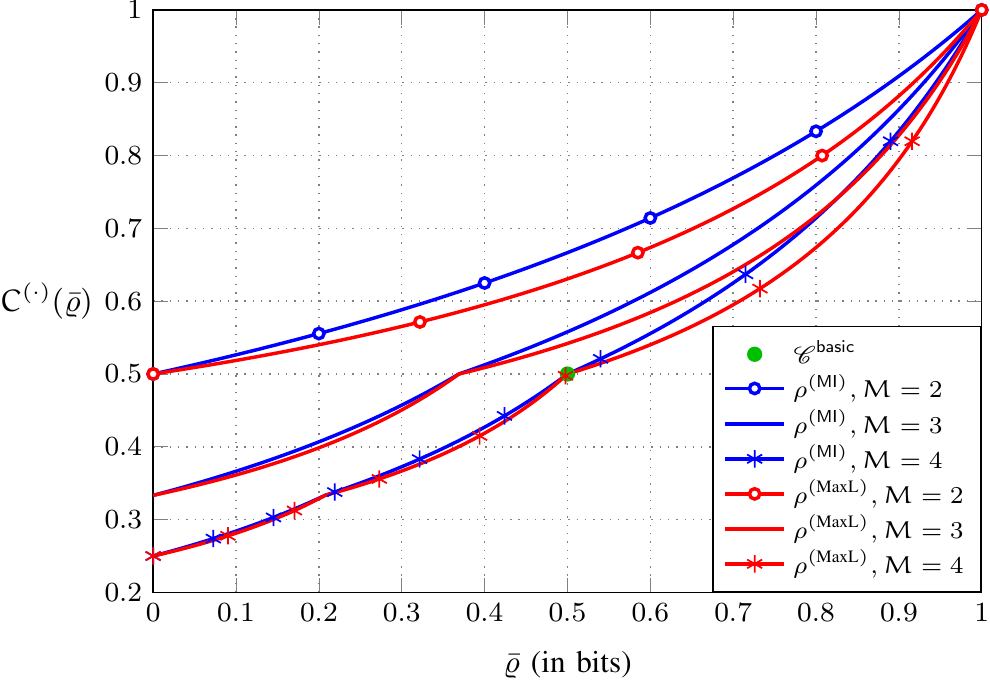}\label{fig:R_SWPIR_M2-4}}
  \hspace*{2mm}
  \subfloat[]{\includegraphics[width=0.5\textwidth]{\Figs/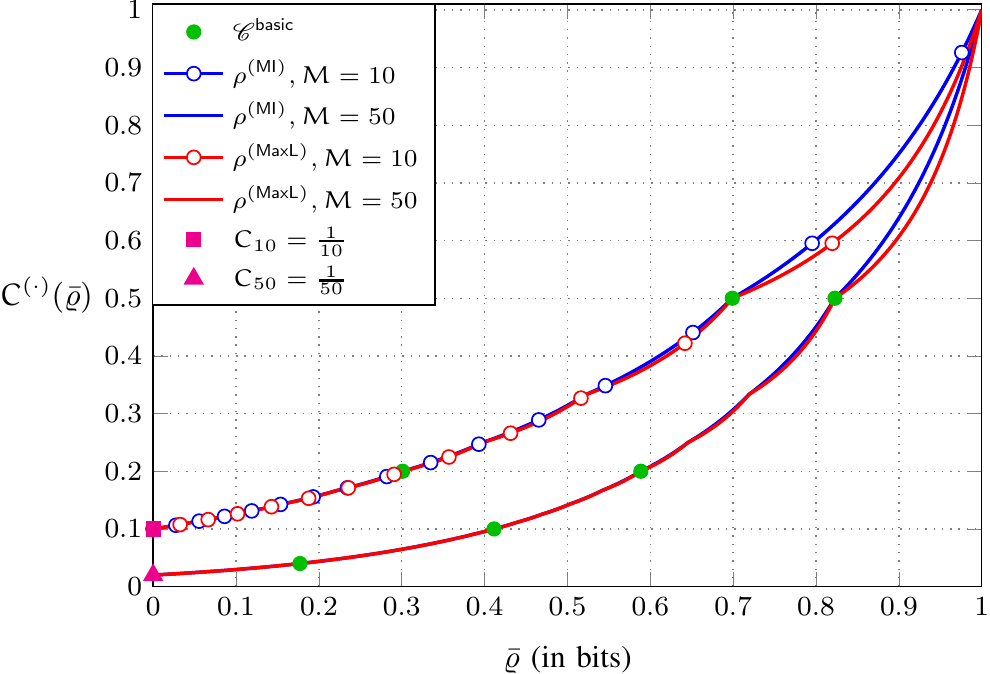}\label{fig:R_SWPIR_M10-50}}
  \caption{(a) The capacity $\const{C}^{(\cdot)}(\bar{\varrho})$ for a small number of files $\const{M}=2,3,4$ with privacy metrics $\rho^{(\mathsf{MI})}$ and $\rho^{(\mathsf{MaxL})}$. (b) The capacity $\const{C}^{(\cdot)}(\bar{\varrho})$ for a large number of files $\const{M}=10,50$ with privacy metrics $\rho^{(\mathsf{MI})}$ and $\rho^{(\mathsf{MaxL})}$. The dark green circles mark the achievable rate-leakage pairs of $\collect{C}^{\mathsf{basic}}$.}
  \label{fig:singleWPIRdwn_M} 
\end{figure*}

The following theorem states the achievable download-leakage pairs of the partition scheme.
\begin{theorem}
  \label{thm:partition-schemes}
  Consider a single server that stores $\const{M}$ files and let $\const{M}_\eta=\const{M}/ \eta \in\Naturals$, $\eta\in\Naturals$. Assume that an $\const{M}_\eta$-file WPIR scheme $\tilde{\collect{C}}$ with achievable download-leakage pair $\bigl(\widetilde{\const{D}}, \tilde{\varrho}\bigr)$ exists. Then, the download-leakage pair
  \begin{IEEEeqnarray}{c}
    \bigl(\const{D}(\collect{C}),\rho^{(\cdot)}(\collect{C})\bigr)=\bigl(\widetilde{\const{D}},\tilde{\varrho}+\log_2{\eta}\bigr)\label{eq:achievable-pair_partition-scheme}
  \end{IEEEeqnarray}
  is achievable by the $\const{M}$-file partition scheme $\collect{C}$ constructed from $\tilde{\collect{C}}$ as described in \eqref{eq:queries_partition-scheme}.
\end{theorem}
\begin{IEEEproof}
  The theorem for the MI privacy metric is proved in~\cite{LinKumarRosnesGraellAmatYaakobi20_2sub} (see proof of Theorem~2). Here, we provide the proof for the MaxL metric, which follows a similar argumentation as in the proof for the MI metric.
  
  We refer to the requested file index $M$ by the pair $(\tilde{M},j)$, where $\tilde{M}$ represents the requested file index in the $j$-th partition, $j\in[\eta]$. Hence, from \eqref{eq:expression_MaxL}, we have
  \begin{IEEEeqnarray}{rCl}
    2^{\ML{M}{\vect{Q}}}& = &\sum_{\vect{q}\in\set{Q}}\max_{m\in[\const{M}]}P_{\vect{Q}|M}(\vect{q}|m)
    \nonumber\\
    & = &\sum_{j\in[\eta]}\sum_{\tilde{\vect{q}}\in\tilde{\set{Q}}}\max_{m\in[\const{M}]}P_{\vect{Q}|M}(\vect{q}|m)
    \nonumber\\
    & \stackrel{(a)}{=} &\sum_{j\in[\eta]}\sum_{\tilde{\vect{q}}\in\tilde{\set{Q}}}\max_{m\in[\const{M}_\eta]}P_{\widetilde{\vect{Q}}|\tilde{M}}(\tilde{\vect{q}}|m)
    \label{eq:use_Mj}\nonumber\\
    & = &\eta\cdot 2^{\ML{\tilde{M}}{\widetilde{\vect{Q}}}}\leq 2^{\log_2{\eta}+\tilde{\varrho}},\nonumber
  \end{IEEEeqnarray}
  where~$(a)$ follows since for the $j$-th partition, the conditional PMF $P_{\vect{Q}|M}$ is equal to $P_{\widetilde{\vect{Q}}|\tilde{M}}$ of the $\const{M}_\eta$-file WPIR scheme. Using a similar argumentation as above, it can also be verified that $\const{D}(\collect{C})=\const{D}(\tilde{\collect{C}})\leq\widetilde{\const{D}}$.
\end{IEEEproof}

Since a PIR scheme is also a WPIR scheme, this simple approach for the construction of WPIR schemes can also be adapted to use any of the existing $\const{M}_{\eta}$-file PIR schemes in the literature as a subscheme. We refer to the partition scheme that uses a PIR scheme as the underlying subscheme and the query generation in~\eqref{eq:queries_partition-scheme} as a \emph{basic scheme} and denote it by $\collect{C}^{\mathsf{basic}}$ (it achieves the pair in~\eqref{eq:achievable-pair_partition-scheme} with $\tilde{\varrho}=0$). It can be seen that for the single-server setting, the basic scheme simply retrieves all the files in the partition that includes the requested file. This idea will be extended to our capacity-achieving scheme presented in Section~\ref{sec:achievability}, where for any subset $\set{M}\subseteq [\const{M}]$ that includes the requested file, all files in $\set{M}$ are downloaded.

\subsection{The Capacity of Single-Server WPIR}
\label{sec:capacity_single-server-WPIR}

The main result of this work is the characterization of the optimal tradeoff between the download cost and the information leakage for single-server WPIR for an arbitrary number of files and a fixed file size $\beta$ for the MI and MaxL privacy metrics. The capacity of single-server WPIR for the MI privacy metric is stated in the following theorem. For the sake of illustration, we consider the normalized leakage metric $\bar{\rho}^{(\cdot)}\eqdef\frac{\rho^{(\cdot)}}{\log_2{\const{M}}}$.

\begin{theorem}
  \label{thm:capacity_SWPIR_MI}
  For a single server that stores $\const{M}$ files, the WPIR capacity for the MI leakage metric $\rho^{(\mathsf{MI})}$ is
  \begin{IEEEeqnarray}{rCl}
    \IEEEeqnarraymulticol{3}{l}{%
      \const{C}^{(\mathsf{MI})}(\bar{\varrho})=\inv{\left[w+\frac{\log_2{\frac{\const{M}}{w}}}{\log_2{\frac{w}{w-1}}}-\frac{\bar{\varrho}\log_2{\const{M}}}{\log_2{\frac{w}{w-1}}}\right]},}
    \nonumber\\*[1mm]
    &&\textnormal{for }1-\frac{\log_2{w}}{\log_2{\const{M}}}\leq\bar{\varrho}\leq 1-\frac{\log_2{(w-1)}}{\log_2{\const{M}}},\,w\in[2:\const{M}].
    \IEEEeqnarraynumspace\label{eq:capacity_SWPIR_MI}
  \end{IEEEeqnarray}   
\end{theorem}

The following theorem states the single-server WPIR capacity for the MaxL privacy metric.
\begin{theorem}
  \label{thm:capacity_SWPIR_MaxL}
    For a single server that stores $\const{M}$ files, 
    the WPIR capacity for the MaxL metric $\rho^{(\mathsf{MaxL})}$ is 
  \begin{IEEEeqnarray*}{rCl}
    \IEEEeqnarraymulticol{3}{l}{%
      \const{C}^{(\mathsf{MaxL})}(\bar{\varrho})=\inv{\left[w+\frac{\frac{\const{M}}{w}}{\frac{\const{M}}{w-1}-\frac{\const{M}}{w}}-\frac{2^{\bar{\varrho}\log_2{\const{M}}}}{\frac{\const{M}}{w-1}-\frac{\const{M}}{w}}\right]},}
    \nonumber\\*[1mm]
    &&\textnormal{for }1-\frac{\log_2{w}}{\log_2{\const{M}}}\leq\bar{\varrho}\leq 1-\frac{\log_2{(w-1)}}{\log_2{\const{M}}},\,w\in[2:\const{M}].\IEEEeqnarraynumspace\label{eq:capacity_SWPIR_MaxL}    
  \end{IEEEeqnarray*}   
\end{theorem}

The achievability proof of Theorems~\ref{thm:capacity_SWPIR_MI} and~\ref{thm:capacity_SWPIR_MaxL} appears in Section~\ref{sec:achievability} while the converse part appears in Sections~\ref{sec:converse_Thm1} and~\ref{sec:converse_Thm2}, for Theorems~\ref{thm:capacity_SWPIR_MI} and~\ref{thm:capacity_SWPIR_MaxL}, respectively.

For the MI and MaxL privacy metrics, the achievable rate-leakage pairs of $\collect{C}^{\mathsf{basic}}$ in~\eqref{eq:achievable-pair_partition-scheme} and the capacity $\const{C}^{(\cdot)}(\bar{\varrho})$ for different number of files $\const{M}$, are depicted in Fig.~\ref{fig:singleWPIRdwn_M}.

  It is worthwhile noting that the curve $\big[\const{C}^{(\cdot)}(\cdot)\big]^{-1}$ is a piecewise continuous function. For the MI privacy metric, \eqref{eq:capacity_SWPIR_MI} indicates that the minimum download cost $\inv{\big[\const{C}^{(\mathsf{MI})}(\cdot)\big]}$ is a piecewise linear function in $\bar{\rho}$, illustrating the convexity of Lemma~\ref{lem:convexity_download-leakage-ft}. Note also that, when $\const{M}_\eta=\const{M} / \eta \in\Naturals$, the basic scheme $\collect{C}^{\mathsf{basic}}$ achieves the capacity for both the MI and MaxL privacy metrics.

Next, we consider the asymptotic capacity of single-server WPIR, i.e., the capacity as the number of files $\const{M}$ tends to infinity. An upper bound on the single-server WPIR capacity for any number of files is given in the following theorem.

\begin{theorem}
  \label{thm:UB_capacity-leakage}
  For a single server that stores $\const{M}$ files,  
  the single-server WPIR capacity under both the MI metric $\rho^{(\mathsf{MI})}$ and the MaxL metric $\rho^{(\mathsf{MaxL})}$ is bounded from above by
  \begin{IEEEeqnarray*}{c}
    \const{C}^{(\cdot)}(\bar{\varrho})\leq\const{C}_{\mathsf{UB}}(\bar{\varrho})\eqdef\frac{1}{\const{M}^{1-\bar{\varrho}}},\quad 0\leq\bar{\varrho}\leq 1.
  \end{IEEEeqnarray*}
\end{theorem}
\begin{IEEEproof}
  Since it is easy to show that the capacity for  MI leakage is larger than or equal to the capacity for MaxL  (see Fig.~\ref{fig:singleWPIRdwn_M}), we only need to prove that the inverse of \eqref{eq:capacity_SWPIR_MI} is bounded from below by $\const{M}^{1-\bar{\varrho}}$. Observe that $\const{M}^{1-\bar{\varrho}}$ is a convex function of $\bar{\varrho}$ and each point $\bigl(w,1-\frac{\log_2{w}}{\log_2{\const{M}}}\bigr)$, $w\in[\const{M}]$, lies on the curve described by that function, i.e., $\const{M}^{1-\bar{\varrho}}=w$ for $\bar{\varrho} = 1-\frac{\log_2{w}}{\log_2{\const{M}}}$. Thus, by the convexity of $\const{M}^{1-\bar{\varrho}}$, we have $\const{M}^{1-\bar{\varrho}}\leq\inv{\bigl[\const{C}^{(\mathsf{MI})}(\bar{\varrho})\bigr]}$, where the inequality follows since the inverse of \eqref{eq:capacity_SWPIR_MI} can be seen as a convex combination of $1-\frac{\log_2{w}}{\log_2{\const{M}}}$ and $1-\frac{\log_2{(w-1)}}{\log_2{\const{M}}}$, $w\in[2:\const{M}]$.
\end{IEEEproof}

\begin{figure}[t!]
  \centering
  \includegraphics[width=0.475\textwidth]{\Figs/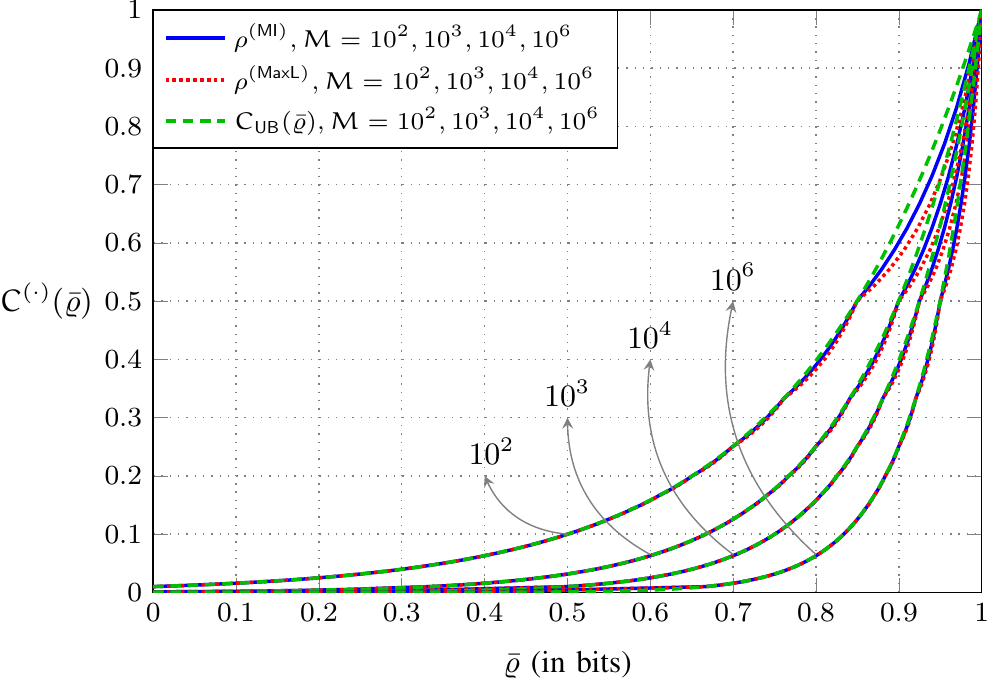}
  \caption{The capacity $\const{C}^{(\cdot)}(\bar{\varrho})$ and its upper bound $\const{C}_\mathsf{UB}(\bar{\varrho})$ for a number of files $\const{M}=10^2$, $10^3$, $10^4$, and $10^6$ with privacy metrics $\rho^{(\mathsf{MI})}$ and $\rho^{(\mathsf{MaxL})}$.}
  \label{fig:singleWPIRdwn_M-huge} 
  \vspace{-2ex} 
\end{figure}

In Fig.~\ref{fig:singleWPIRdwn_M-huge}, the capacity $\const{C}^{(\cdot)}(\bar{\varrho})$ and the upper bound $\const{C}_\mathsf{UB}(\bar{\varrho})$ are plotted for $\const{M}=10^2$, $10^3$, $10^4$, and $10^6$, which illustrates the asymptotic behavior of $\const{C}^{(\cdot)}(\bar{\varrho})$ as $\const{M}$ tends to infinity. Note that since one can simply download the requested file from the server in the special case of $\bar{\rho}=1$, and the WPIR rate must be smaller than or equal to $1$, 
we have $\const{C}^{(\cdot)}(1)=1$ for either a finite or infinite number of files $\const{M}$. Hence, by Theorem~\ref{thm:UB_capacity-leakage} it can be shown that as $\const{M}$ tends to infinity, the asymptotic capacity is equal to
\begin{IEEEeqnarray*}{c}
  \const{C}^{(\cdot)}_{\infty}(\bar{\varrho})=
  \begin{cases}
    0 &\textnormal{if }0\leq\bar{\varrho}<1,
    \\
    1 &\textnormal{if }\bar{\varrho}=1.
  \end{cases}
\end{IEEEeqnarray*}
This indicates that the asymptotic capacity is still equal to zero, unless the server exactly knows the index of the requested file.

\section{Achievability}
\label{sec:achievability}

Throughout this section, 
for simplicity, we set $\beta=1$. Hence, from Section~\ref{sec:system-model}, we have $\vect{X}^{(m)}=X_1^{(m)}$ and $\eHP{\vect{X}^{(m)}}=\log_2{\card{\set{X}}}$ bits, $\forall\,m\in[\const{M}]$. In fact, our proposed single-server WPIR capacity-achieving scheme can be easily generalized to an arbitrary file size $\beta$, which indicates that subpacketization does not improve the performance of single-server WPIR. We will later show that this scheme is optimal for both the MI and MaxL privacy metrics.

\subsection{Motivating Example: $\const{M}=3$ Files}
\label{sec:SWPIR_M3}

Before describing the achievable scheme in detail for the general case of $\const{M}$ files, we present an example for  $\const{M}=3$. 
Assume that the single server stores $\const{M}=3$ files, $X^{(1)}_1$, $X^{(2)}_1$, and $X^{(3)}_1$. We design the queries and answers via a conditional distribution $P_{\vect{Q}_w|M}$, $w\in[3]$, defined in Table~\ref{tab:achvPMFs_M3}. It can be easily verified that the perfect retrievability condition of \eqref{eq:retrievability} is satisfied for the three PMFs. 
Moreover, the download-leakage pairs $(\const{D}_w,\varrho_w)=(w,\log_2{\frac{3}{w}})$ are achievable, by $P_{\vect{Q}_w|M}$, $w\in[3]$, in terms of the MI or MaxL privacy metrics. 

\begin{table}[t!]
  \centering
  \caption{The conditional PMFs $P_{\vect{Q}_w|M}$, $w\in[3]$.}
  \label{tab:achvPMFs_M3}  
   \vskip -2.0ex %
  \Resize[0.95\columnwidth]{
    \begin{IEEEeqnarraybox}[
      \IEEEeqnarraystrutmode
      \IEEEeqnarraystrutsizeadd{6pt}{5pt}]{v/c/V/c/v/c/v/c/v/c/v/c/v}
      \IEEEeqnarrayrulerow\\
      &\vect{Q}_1 && P_{\vect{Q}_1|M}(\vect{q}\mid 1)  && P_{\vect{Q}_1|M}(\vect{q}\mid 2) && P_{\vect{Q}_1|M}(\vect{q}\mid 3) && \vect{A}_1 && P_{\vect{Q}_1}(\vect{q})
      \\\hline\hline
      & (1,0,0) &&1 && 0 && 0 && X^{(1)}_1 && \frac{1}{3}
      \\*\hline
      & (0,1,0) && 0&& 1 && 0 && X^{(2)}_1 && \frac{1}{3}
      \\*\hline
      & (0,0,1) && 0&& 0 && 1 && X^{(3)}_1 && \frac{1}{3}
      \\*\IEEEeqnarrayrulerow
    \end{IEEEeqnarraybox}}
  \\[3mm]
  \Resize[0.95\columnwidth]{
    \begin{IEEEeqnarraybox}[
      \IEEEeqnarraystrutmode
      \IEEEeqnarraystrutsizeadd{6pt}{5pt}]{v/c/V/c/v/c/v/c/v/c/v/c/v}
      \IEEEeqnarrayrulerow\\
      &\vect{Q}_2 && P_{\vect{Q}_2|M}(\vect{q}\mid 1) && P_{\vect{Q}_2|M}(\vect{q}\mid 2) && P_{\vect{Q}_2|M}(\vect{q}\mid 3) && \vect{A}_2 && P_{\vect{Q}_2}(\vect{q})
      \\\hline\hline
      & (1,1,0) && \frac{1}{2} && \frac{1}{2} && 0 && \bigl\{X^{(1)}_1,X^{(2)}_1\bigr\} && \frac{1}{3}
      \\*\hline
      & (1,0,1) && \frac{1}{2} && 0 && \frac{1}{2} && \bigl\{X^{(1)}_1,X^{(3)}_1\bigr\} && \frac{1}{3}
      \\*\hline
      & (0,1,1) && 0 && \frac{1}{2} && \frac{1}{2} && \bigl\{X^{(2)}_1,X^{(3)}_1\bigr\} && \frac{1}{3}
      \\*\IEEEeqnarrayrulerow
    \end{IEEEeqnarraybox}}
  \\[3mm]
    \Resize[0.95\columnwidth]{
    \begin{IEEEeqnarraybox}[
      \IEEEeqnarraystrutmode
      \IEEEeqnarraystrutsizeadd{6pt}{5pt}]{v/c/V/c/v/c/v/c/v/c/v/c/v}
      \IEEEeqnarrayrulerow\\
      &\vect{Q}_3 && P_{\vect{Q}_3|M}(\vect{q}\mid 1) && P_{\vect{Q}_3|M}(\vect{q}\mid 2) && P_{\vect{Q}_3|M}(\vect{q}\mid 3) && \vect{A}_3 && P_{\vect{Q}_3}(\vect{q})
      \\\hline\hline
      & (1,1,1) && 1 && 1 && 1 && \bigl\{X^{(1)}_1,X^{(2)}_1,X^{(3)}_1\bigr\} && 1
      \\*\IEEEeqnarrayrulerow
    \end{IEEEeqnarraybox}}
\end{table}

Now, construct two conditional query distributions as follows,
\begin{IEEEeqnarray}{rCl}
  P_{\vect{Q}_{\lambda_1}|M}& = &(1-\lambda_1)P_{\vect{Q}_{2}|M}+\lambda_1 P_{\vect{Q}_{1}|M},
  \label{eq:case1}
  \\[1mm]
  P_{\vect{Q}_{\lambda_2}|M}& = &(1-\lambda_2)P_{\vect{Q}_{3}|M}+\lambda_2 P_{\vect{Q}_{2}|M},
  \label{eq:case2}
\end{IEEEeqnarray}
where $0\leq\lambda_1,\lambda_2\leq 1$. 

The retrievability condition can be easily verified for the WPIR scheme defined by~\eqref{eq:case1}--\eqref{eq:case2}. 
Using~\eqref{eq:D-def_single-serverWPIR}, \eqref{eq:expression_MI}, and \eqref{eq:expression_MaxL}, respectively, and the conditional PMFs listed in Table~\ref{tab:achvPMFs_M3} (or, alternatively, Corollary~\ref{cor:convexity_achievable-pairs}), it follows that the WPIR scheme defined by \eqref{eq:case1}--\eqref{eq:case2} achieves the download cost  
\begin{IEEEeqnarray*}{c}
  \const{D}(\collect{C})=
  \begin{cases}
  (1-\lambda_1)\const{D}_{2}+\lambda_1\const{D}_{1}=2-\lambda_1, &\; 0\leq\lambda_1\leq 1,
  \\[1mm]
  (1-\lambda_2)\const{D}_{3}+\lambda_2\const{D}_{2}=3-\lambda_2, &\; 0\leq\lambda_2\leq 1,
  \end{cases}
\end{IEEEeqnarray*}
the MI leakage
\begin{IEEEeqnarray*}{c}
  \rho^{(\mathsf{MI})}=
  \begin{cases}
    (1-\lambda_1)\log_2{\frac{3}{2}}+\lambda_1\log_2{\frac{3}{1}}, & 0\leq\lambda_1\leq 1,
    \\[1mm]
    (1-\lambda_2)\log_2{\frac{3}{3}}+\lambda_2\log_2{\frac{3}{2}}, & 0\leq\lambda_2\leq 1,
  \end{cases}
\end{IEEEeqnarray*}
and the MaxL
\begin{IEEEeqnarray*}{c}
  \rho^{(\mathsf{MaxL})}=
  \begin{cases}
    \log_2\bigl((1-\lambda_1)\frac{3}{2}+\lambda_1\frac{3}{1}\bigr), & 0\leq\lambda_1\leq 1,
    \\[1mm]
    \log_2\bigl((1-\lambda_2)\frac{3}{3}+\lambda_2\frac{3}{2}\bigr), & 0\leq\lambda_2\leq 1.
  \end{cases}
\end{IEEEeqnarray*}
In terms of the MI or MaxL privacy metrics, it can  be verified that the download cost corresponds to the single-server WPIR capacity for $\const{M}=3$. Note that for $\const{M}>2$, the capacity is a piecewise continuous function (see Fig.~\subref*{fig:R_SWPIR_M2-4}).

\subsection{Arbitrary Number of Files $\const{M}$}
\label{sec:general-M}

We describe the achievable scheme for the general case of $\const{M}$ files. From Corollary~\ref{cor:convexity_achievable-pairs}, it follows that it is sufficient to show that the download-leakage pairs 
\begin{IEEEeqnarray*}{c}
  (\const{D}_w,\varrho_w)=\left(w,\log_2{\frac{\const{M}}{w}}\right),\quad w\in[\const{M}],
\end{IEEEeqnarray*}
are achievable.

\subsubsection{Query Generation}
\label{sec:query-generation_SWPIR}

Consider $\const{M}$ random queries $\vect{Q}_w$, $w\in[\const{M}]$, whose alphabet is $\set{Q}_w\eqdef\{\vect{q}=(q_1,\ldots,q_\const{M})\in\{0,1\}^{\const{M}}:\Hwt{\vect{q}}=w\}$. Recall here that $\Spt{\vect{q}}$ denotes the support of a vector $\vect{q}$. Given any requested file index $m\in[\const{M}]$, each query $\vect{q}\in\set{Q}_w$ sent to the server is generated by the conditional PMF
\begin{IEEEeqnarray*}{c}
  P_{\vect{Q}_w|M}(\vect{q}|m)=
  \begin{cases}
    \frac{1}{\binom{\const{M}-1}{w-1}} & \textnormal{if } \bigcard{\chi{(\vect{q})}\setminus\{m\}}=w-1,
    \\
    0 & \textnormal{otherwise}.
  \end{cases}
\end{IEEEeqnarray*}
This is clearly a valid query design, since for each $m\in[\const{M}]$, we have $\sum_{\vect{q}\in\set{Q}_w}P_{\vect{Q}_w|M}(\vect{q}|m)
=1$.

\subsubsection{Answer Construction}
\label{sec:answer-construction_SWPIR}

The answer function $\varphi$ maps the query $\vect{q}\in\set{Q}_w$ onto $\vect{A}=\varphi(\vect{q},\vect{X}^{[\const{M}]})=X_1^{\chi(\vect{q})}$. The answer length is $L(\vect{q})=w$.

\subsubsection{Download Cost and Information Leakage}
\label{sec:download-leakage_SWPIR}

Clearly, the download cost is equal to $\E[P_{M}P_{\vect{Q}_w|M}]{L(\vect{Q}_w)}=w$. The MI leakage is
\begin{IEEEeqnarray*}{rCl}
  \rho^{(\mathsf{MI})}(P_{\vect{Q}_w|M})& = &\MI{M}{\vect{Q}_w}=\eHP{M}-\eHPcond{M}{\vect{Q}_w}
  \\
  & = &\log_2{\const{M}}-\log_2{w}=\log_2{\frac{\const{M}}{w}}
\end{IEEEeqnarray*}
and the MaxL is
\begin{IEEEeqnarray*}{rCl}
  \rho^{(\mathsf{MaxL})}(P_{\vect{Q}_w|M})& = &\log_2{\sum_{\vect{q}\in\set{Q}_w}\max_{m\in[\const{M}]}\frac{1}{\binom{\const{M}-1}{w-1}}}
  \\[1mm]
  & = &\log_2{\frac{\binom{\const{M}}{w}}{\binom{\const{M}-1}{w-1}}}=\log_2{\frac{\const{M}}{w}}.
\end{IEEEeqnarray*}
This completes the achievability proof of Theorems~\ref{thm:capacity_SWPIR_MI} and \ref{thm:capacity_SWPIR_MaxL}.

Notice that the presented capacity-achieving WPIR scheme can be seen as a generalization of the basic WPIR scheme $\collect{C}^{\mathsf{basic}}$. If $\const{M} / \eta=w\in\Naturals$, since $(\widetilde{\const{D}},\tilde{\varrho})=(w,0)$ is achievable for a $w$-file single-server PIR scheme, from Theorem~\ref{thm:partition-schemes} it follows that the download-leakage pair $\bigl(\const{D}(\collect{C}^{\mathsf{basic}}),\rho^{(\cdot)}(\collect{C}^{(\mathsf{basic})})\bigr)=(w,\log_2{\eta})=(w,\log_2{\frac{\const{M}}{w}})$ is also achievable by $\collect{C}^{\mathsf{basic}}$ for both the MI and MaxL privacy metrics.

\section{Converse of Theorem~\ref{thm:capacity_SWPIR_MI}}
\label{sec:converse_Thm1}

For any file size $\beta$, a general converse (upper bound) 
can be derived from the download-leakage function of a given leakage constraint $\varrho$, or equivalently, from the leakage-download function of a given download cost constraint $\const{D}$. The proof consists of two parts, and we start by outlining the main arguments of each part before diving further into the technical details. 


\subsubsection*{Part~1}

Consider an arbitrary WPIR scheme. Without loss of optimality, for any query sent to the server, the answer from the server can be assumed to be a subset of the files that includes the desired file. This is because in order to have perfect retrievability, downloading any linear combinations, or any coded forms of a subset of the files can only lead to a higher download cost. Moreover, it can also not increase the privacy leakage, because the server can only infer the identity of the desired file from an answer that is able to recover a subset of the files. We formally prove this argument later in this section. Note that this part is also used in the converse proof of Theorem~\ref{thm:capacity_SWPIR_MaxL} as it holds for both the MI and MaxL privacy metrics.
%
%
%
%

\subsubsection*{Part~2} 
From Part~1 we can limit the consideration to schemes for which all answers are subsets of files that include the desired file.
Since the minimum achievable information leakage for a given download cost constraint among this limited family of schemes can be related to the rate-distortion function with a certain distortion measure, we can apply a known lower bound on the  rate-distortion function in order to find an optimal scheme from this family. Finally, we show that the optimal scheme is exactly the scheme we propose in Section~\ref{sec:achievability}. Thus, for a given download cost constraint, the leakage of any WPIR scheme is bounded below by the leakage of the scheme proposed in Section~\ref{sec:achievability}.

We start to prove the first part of the converse proof. Given an arbitrary query set $\set{Q}$ of a WPIR scheme with $\{L(\vect{q})\}_{\vect{q}\in\set{Q}}\subseteq\set{L}_{\mathsf{ret}}$, similar to~\eqref{eq:download-leakage-ft}, the minimum leakage of this WPIR scheme for a given download cost constraint $\const{D}$ can be formulated as the convex optimization problem 
\begin{IEEEeqnarray}{rCl}
  \IEEEyesnumber\label{eq:leakage-download_MI}
  \IEEEyessubnumber*
  \textnormal{minimize} & &\qquad\MI{M}{\vect{Q}}
  \\
  \textnormal{subject to} & &\qquad\E[P_{M}P_{\vect{Q}|M}]{L(\vect{Q})}\leq\const{D}. \label{eq:PMFs_download-constraint}
\end{IEEEeqnarray}

The minimization is taken over the set of all conditional distributions $P_{\vect{Q}|M}$ such that \eqref{eq:PMFs_download-constraint} is satisfied, namely the set
\begin{IEEEeqnarray*}{c}
  \set{F}_{\const{D}}=\Biggl\{P_{\vect{Q}|M}\colon
  \!\!\!\sum_{\vect{q}}\!\!\sum_{m\in[\const{M}]}P_M(m)P_{\vect{Q}|M}(\vect{q}|m)L(m,\vect{q})\leq\const{D}\Biggr\}.\IEEEeqnarraynumspace
\end{IEEEeqnarray*}

Next, we show a lower bound to \eqref{eq:leakage-download_MI} by defining a new RV  that is a function of $\vect{Q}$. To facilitate the exposition, we introduce the following notation.
\begin{itemize}
\item For any nonempty subset $\set{M}\subseteq[\const{M}]$, we define $\tilde{\set{Q}}^{\set{M}}$ to be the set of queries that are designed to recover the files $\vect{X}^{(m)}$, $m\in\set{M}$, i.e., $\tilde{\set{Q}}^{\set{M}}\eqdef\bigl\{\vect{q}\in\set{Q}\colon\eHPcond{\vect{X}^{(m)}}{\vect{A},\vect{Q}=\vect{q},M=m,\vect{S}=\vect{s}}=0,\,\forall\,m\in\set{M}\bigr\}$. Furthermore, define 
  \begin{IEEEeqnarray*}{c}
      \set{Q}^{\set{M}}\eqdef\tilde{\set{Q}}^{\set{M}}\setminus\Biggl(\bigcup_{\set{M}'\subseteq[\const{M}]\setminus\set{M}}\tilde{\set{Q}}^{\set{M}'}\Biggr).
  \end{IEEEeqnarray*}
  The set $\set{Q}^{\set{M}}$ contains all queries that are designed to recover all files in $\set{M}$, but no more. Note that if $\vect{q}\in\set{Q}^{\set{M}}$ but $m\notin\set{M}$, then $P_{\vect{Q}|M}(\vect{q}|m)=0$.
  
\item A binary length-$\const{M}$ \emph{indicator vector} $\vect{1}_{\set{M}}=(u_1,\ldots,u_\const{M})$ of a subset $\set{M}\subseteq [\const{M}]$, $\set{M}\neq\emptyset$, is defined as
\begin{IEEEeqnarray*}{c}
  u_m=
  \begin{cases}
    1 & \textnormal{if } m\in\set{M},
    \\
    0 & \textnormal{otherwise}.
  \end{cases}
\end{IEEEeqnarray*}
\end{itemize}

$\vect{Q}$ is a RV that is induced by the conditional distribution $P_{\vect{Q}|M}$. Further define a new RV $\vect{U}=f(\vect{Q})$, where
\begin{IEEEeqnarray}{c} \label{eq:U}
  f(\vect{q})\eqdef\vect{1}_{\set{M}},\quad\textnormal{for }\vect{q}\in\set{Q}^{\set{M}},\,\forall\,\set{M}\subseteq[\const{M}],\,\set{M}\neq\emptyset.
\end{IEEEeqnarray}
Note that the mapping in \eqref{eq:U} is well-defined since the query sets $\set{Q}^{\set{M}}$ are disjoint and their union is equal $\set{Q}$, i.e., they constitute a partition of $\set{Q}$. This leads to the conditional PMF 
\begin{IEEEeqnarray}{rCl}
  P_{\vect{U}|M}(\vect{u}|m)=
  \begin{cases}
    \sum_{\vect{q}\in\set{Q}^{\chi(\vect{u})}}P_{\vect{Q}|M}(\vect{q}|m) & \textnormal{if }m\in\chi(\vect{u}),
    \\[1mm]
    0 & \textnormal{otherwise}.
  \end{cases}
  \IEEEeqnarraynumspace\label{eq:definition_PU}
\end{IEEEeqnarray}

To further simplify the notation, in the following we use $p(\vect{q}|m)$ and $p(\vect{u}|m)$ to denote the conditional PMFs $P_{\vect{Q}|M}$ and $P_{\vect{U}|M}$, respectively.

We then design a WPIR scheme where the queries are generated according to $p(\vect{u}|m)$ and the normalized answer-length function is constructed for any $\vect{u}\in\{0,1\}^\const{M}$ as
\begin{IEEEeqnarray*}{c}
  L(m,\vect{u})=
  \begin{cases}
    \Hwt{\vect{u}} & \textnormal{if } m\in\chi(\vect{u}),
    \\
    \infty & \textnormal{otherwise},
  \end{cases}
\end{IEEEeqnarray*}
where an infinite length for a given file index $m$ and query $\vect{u}$ indicates that the $m$-th file is never retrieved by the designed query $\vect{u}$. 

Accordingly, we define the set
\begin{IEEEeqnarray*}{rCl}
  \set{P}_{\const{D}}& = &\biggl\{p(\vect{q}|m)\colon\sum_{\vect{u}\in\{0,1\}^{\const{M}}}\sum_{\vect{q}\in\set{Q}^{\chi(\vect{u})}}\sum_{m\in[\const{M}]}\nonumber\\
  &&\hspace{2.5cm}
  P_M(m)p(\vect{q}|m)L(m,\vect{u})\leq\const{D}\biggr\}.\IEEEeqnarraynumspace
\end{IEEEeqnarray*}
Given an arbitrary $p(\vect{q}|m)\in\set{F}_{\const{D}}$, 
\begin{IEEEeqnarray}{rCl}
  \IEEEeqnarraymulticol{3}{l}{%
    \sum_{\vect{u}\in\{0,1\}^{\const{M}}}\sum_{m\in[\const{M}]}P_M(m)p(\vect{u}|m)L(m,\vect{u})}\nonumber\\*\,%
  & = &\sum_{\vect{u}\in\{0,1\}^{\const{M}}}\sum_{m\in[\const{M}]}P_M(m)\sum_{\vect{q}\in\set{Q}^{\chi(\vect{u})}}p(\vect{q}|m)L(m,\vect{u})
  \nonumber\\
  & \stackrel{(a)}{\leq} &\sum_{\vect{u}\in\{0,1\}^{\const{M}}}\sum_{\vect{q}\in\set{Q}^{\chi(\vect{u})}}\sum_{m\in[\const{M}]}P_M(m)p(\vect{q}|m)
  L(m,\vect{q})\label{eq:use_Q-length}\IEEEeqnarraynumspace\nonumber
  \\
  & = &
  \sum_{\vect{q}\in\set{Q}}\sum_{m\in[\const{M}]}P_M(m)p(\vect{q}|m)L(m,\vect{q})\leq\const{D},\nonumber
\end{IEEEeqnarray}
where $(a)$ holds because for any $\vect{q}\in\set{Q}^{\chi(\vect{u})}$ we know from the lossless source coding theorem that the normalized length of the answer $L(m,\vect{q})$ is larger than or equal to the ratio between the total sizes of the retrieved $\Hwt{\vect{u}}=L(m,\vect{u})$ files and the logarithm of the source code's alphabet, i.e., it satisfies $L(m,\vect{q})\geq \nicefrac{L(m,\vect{u})\log_2{\card{\set{X}}}}{\log_2{\card{\set{A}}}}=L(m,\vect{u})$.\footnote{The fundamental theorem of lossless data compression states that the expected codeword length is no less than $\nicefrac{\HP{\vect{X}^{\Spt{\vect{u}}}}}{\log_2{\card{\set{A}}}}=\nicefrac{\Hwt{\vect{u}}\log_2{\card{\set{X}}}}{\log_2{\card{\set{A}}}} = L(m,\vect{u})$, and the minimal expected codeword length can be achieved by an optimal source code, e.g., a Huffman code (cf.~\cite[Th.~5.4.1]{CoverThomas06_1}). Here, since the files/sources are i.i.d.~according to a uniform distribution, the codeword lengths are identical.} Hence, it follows that $p(\vect{q}|m)$ also lies in $\set{P}_{\const{D}}$, and hence $\set{F}_{\const{D}}\subseteq\set{P}_{\const{D}}$. The intuition behind this fact is that the function $L(m,\vect{u})$ defines the minimum required lengths of answers for a valid single-server WPIR scheme, thus we have more choices of conditional PMFs $p(\vect{q}|m)$ in $\set{P}_{\const{D}}$.

Now, if we take the minimization over $\set{P}_{\const{D}}$, which is a candidate set that is larger than $\set{F}_{\const{D}}$, we have
\begin{IEEEeqnarray}{rCl}
    \min_{p(\vect{q}|m)\in\set{F}_{\const{D}}}\MI{M}{\vect{Q}}  
  & \geq &\min_{p(\vect{q}|m)\in\set{P}_{\const{D}}}\MI{M}{\vect{Q}}
  \nonumber\\*[1mm]\qquad%
  & \stackrel{(a)}{\geq} &\min_{p(\vect{u}|m)\colon\E[]{L(M,\vect{U})}\leq\const{D}}\MI{M}{\vect{U}},
  \IEEEeqnarraynumspace\label{eq:minimization_PU-M}
\end{IEEEeqnarray}
where $(a)$ follows directly from the data processing inequality of  the first statement of Lemma~\ref{lem:convex_MI-MaxL}. Therefore, a lower bound to the convex optimization problem~\eqref{eq:leakage-download_MI} is given.

In the second part of the proof, we show that \eqref{eq:minimization_PU-M} admits a closed-form expression by using a useful result from rate-distortion theory, a lower bound on the rate-distortion function.\footnote{The proof of this lower bound is based on the Karush–Kuhn–Tucker optimality conditions, see the details in~\cite[Ch.~9]{Gallager68_1}, \cite[Ch.~10]{moser19_v43}.} This lower bound  is adapted to~\eqref{eq:minimization_PU-M} and is re-stated as follows.
\begin{lemma}[{\cite[Th.~9.4.1]{Gallager68_1},~\cite[Th.~10.19]{moser19_v43}}]
  \label{lem:lower-bound_rate-distortion-ft}
  Given any $\const{D}\in[\const{M}]$, we have
  \begin{IEEEeqnarray}{rCl}
    \IEEEeqnarraymulticol{3}{l}{%
      \min_{p(\vect{u}|m)\colon\E[]{L(M,\vect{U})}\leq\const{D}}\MI{M}{\vect{U}}
    }\nonumber\\*\qquad\qquad\qquad%
    & \geq &\eHP{M}+\sum_{m\in[\const{M}]}P_M(m)\log_2{\nu_m}-\lambda\const{D}
    \IEEEeqnarraynumspace\label{eq:lower-bound_RD}
  \end{IEEEeqnarray}
  for an arbitrary choice of $\lambda>0$ and for any $\nu_m$, $m\in[\const{M}]$, satisfying
  \begin{IEEEeqnarray}{c}
    \sum_{m\in [\const{M}]}\nu_m 2^{-\lambda L(m,\vect{u})}\leq 1,\quad\vect{u}\in\{0,1\}^{\const{M}}.
    \label{eq:conditions_lower-bound-RD}
  \end{IEEEeqnarray}
\end{lemma}
We remark again that the length function $L(m,\vect{u})$ is equal to $\Hwt{\vect{u}}$ for all $m\in\chi(\vect{u})$. By using the condition in  \eqref{eq:conditions_lower-bound-RD} for each $\vect{u}$ with $\Hwt{\vect{u}}=1$, we obtain
\begin{IEEEeqnarray*}{rCl}
  \sum_{m\in[\const{M}]}\nu_m 2^{-\lambda L(m,\vect{u})}& = &\sum_{m\in\chi(\vect{u})}\nu_m 2^{-\lambda \Hwt{\vect{u}}}
  \\[1mm]
  & = &\nu_{m}2^{-\lambda}\leq 1,\quad\textnormal{where } m\in\chi(\vect{u}).
\end{IEEEeqnarray*}
This implies that for any $m\in[\const{M}]$, $\nu_m\cdot 2^{-\lambda}\leq 1$, and hence by symmetry, we can simply assume that
\begin{IEEEeqnarray*}{c}
  \nu_m=\nu,\quad\forall\,m.  
\end{IEEEeqnarray*}

Next, we apply \eqref{eq:conditions_lower-bound-RD} for all $\vect{u}\in\{0,1\}^{\const{M}}$:
\begin{IEEEeqnarray*}{rCl}
  \nu\cdot 2^{-1\lambda}& \leq &1\quad\textnormal{if }\Hwt{\vect{u}}=1,
  \\
  2\nu\cdot 2^{-2\lambda}& \leq &1\quad\textnormal{if }\Hwt{\vect{u}}=2,
  \\
  3\nu\cdot 2^{-3\lambda}& \leq &1\quad\textnormal{if }\Hwt{\vect{u}}=3,
  \\
  & & \vdots
  \\
  \const{M}\nu\cdot 2^{-\const{M}\lambda}& \leq &1\quad\textnormal{if }\Hwt{\vect{u}}=\const{M}.
\end{IEEEeqnarray*}
From the above conditions, we obtain
\begin{IEEEeqnarray*}{rCl}
  \log_2{\nu}& \leq &
  \begin{cases}
    \lambda& \textnormal{if } \lambda>\log_2{\frac{2}{1}},
    \\
    2\lambda-\log_2{2}& \textnormal{if } \log_2{\frac{3}{2}}<\lambda\leq\log_2{\frac{2}{1}},
    \\
    3\lambda-\log_2{3}& \textnormal{if } \log_2{\frac{4}{3}}<\lambda\leq\log_2{\frac{3}{2}},
    \\
    & \vdots
    \\
    \const{M}\lambda-\log_2{\const{M}}& \textnormal{if } 0<\lambda\leq\log_2{\frac{\const{M}}{\const{M}-1}}.
  \end{cases}
\end{IEEEeqnarray*}
Now, taking
\begin{IEEEeqnarray*}{rCl}
  (\log_2{\nu},\lambda)=\Bigl(w\lambda-\log_2{w},\log_2{\frac{w}{w-1}}\Bigr),\quad w\in [2:\const{M}],
\end{IEEEeqnarray*}
and substituting this in \eqref{eq:lower-bound_RD} with $P_M(m)=\frac{1}{\const{M}}$, we have 
\begin{IEEEeqnarray}{rCl}
  \IEEEeqnarraymulticol{3}{l}{%
    \min_{p(\vect{u}|m)\colon\E[]{L(M,\vect{U})}\leq\const{D}}\MI{M}{\vect{U}}
  }\nonumber\\*\quad%
  & \geq &\log_2{\const{M}}+\Bigl(w\log_2{\frac{w}{w-1}}-\log_2{w}\Bigr)
  -\Bigl(\log_2{\frac{w}{w-1}}\Bigr)\const{D}
  \nonumber\\[1mm]
  & = &\log_2{\frac{\const{M}}{w-1}}-\Bigl(\log_2{\frac{w}{w-1}}\Bigr)(\const{D}-(w-1)),
  \IEEEeqnarraynumspace\label{eq:privacy_D}
\end{IEEEeqnarray}
for $w\in[2:\const{M}]$.

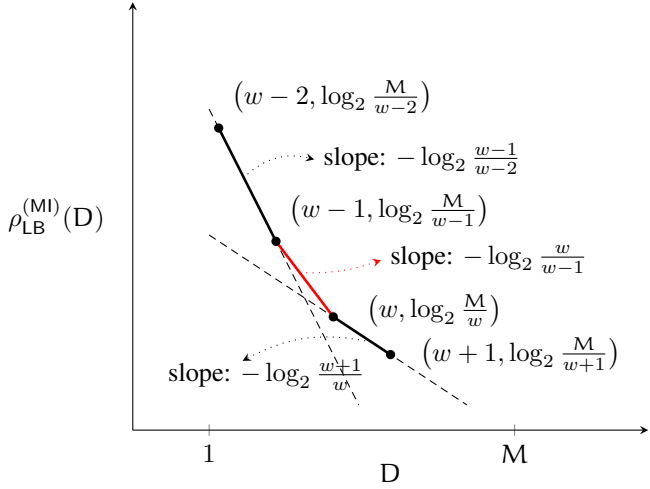
\begin{figure}[t!]
\centering
\begin{tikzpicture}[scale=1, every node/.style={transform shape}, thick]
  \begin{axis}[
    axis x line=bottom,
    axis y line=left,
    xmin=-3, xmax=24, 
    ymin=-1, ymax=16,
    xlabel={$\const{D}$},
    xlabel style={yshift=1.5ex,name=label},
    every axis y label/.style={at={(current axis.west)},left=2.5mm},
    ylabel={$\rho_\mathsf{LB}^{(\mathsf{MI})}(\const{D})$},
    ytick=\empty,
    xticklabels={$1$,$\const{M}$},
    xtick={1,17},
    ]
    \draw[solid,line width=1pt] (axis cs:1.5,11) to (axis cs:4.5,6.5);
    \draw[>=stealth,->,dotted] (axis cs:2.77,9.0) to [bend left=30] (axis cs:6.5,9.75);
    \node at (axis cs:6.5,9.75) [right] {slope: $-\log_2{\frac{w-1}{w-2}}$};
    \draw[densely dashed] (axis cs:1,11.75) to (axis cs:1.5+22/3,0); 
    \draw[red,solid,line width=1pt] (axis cs:4.5,6.5) to (axis cs:7.5,3.5);
    \draw[red,>=stealth,->,dotted] (axis cs:5.75,5.25) to [bend right=10] (axis cs:10.0,5.75);
    \node at (axis cs:10.0,5.75) [right] {slope: $-\log_2{\frac{w}{w-1}}$};
    \draw[solid,line width=1pt] (axis cs:7.5,3.5) to (axis cs:10.5,2);
    \draw[densely dashed] (axis cs:1,6.75) to (axis cs:14.5,0); 
    \draw[>=stealth,->,dotted] (axis cs:9.5,2.5) to [bend right=20] (axis cs:2.75,1.75);
    \node at (axis cs:4.0,1.5) [yshift=-1mm] {slope: $-\log_2{\frac{w+1}{w}}$};
    \node[circle, fill=black, inner sep=0pt, minimum size=3.5pt,label=above right:{$\bigl(w-2,\log_2{\frac{\const{M}}{w-2}}\bigr)$}] at (axis cs:1.5,11) {};
    \node[circle, fill=black, inner sep=0pt, minimum size=3.5pt,label=above right:{$\bigl(w-1,\log_2{\frac{\const{M}}{w-1}}\bigr)$}] at (axis cs:4.5,6.5) {};
    \node[circle, fill=black, inner sep=0pt, minimum size=3.5pt,label={[above right = -0.25 cm and 0.25 cm]0:$\bigl(w,\log_2{\frac{\const{M}}{w}}\bigr)$}] at (axis cs:7.5,3.5) {};
    \node[circle, fill=black, inner sep=0pt, minimum size=3.5pt,label={[label distance=0.2cm]0:$\bigl(w+1,\log_2{\frac{\const{M}}{w+1}}\bigr)$}] at (axis cs:10.5,2) {};    
  \end{axis}
\end{tikzpicture}
\caption{An illustration of the function $\rho_\mathsf{LB}^{(\mathsf{MI})}(\const{D})$, which is defined by many linear functions. The slope of the linear functions is strictly increasing in $w\in[2:\const{M}]$.}
\label{Fig:rho_D}
\end{figure}

Here, \eqref{eq:privacy_D} is a linear function of $\const{D}$ with slope $-\lambda=-\log_2{\frac{w}{w-1}}$, which is strictly increasing in $w\in[2:\const{M}]$. Therefore, the best lower bound for \eqref{eq:minimization_PU-M} 
is the piecewise function 
\begin{IEEEeqnarray*}{rCl}
  \rho_\mathsf{LB}^{(\mathsf{MI})}(\const{D})& = &
  \begin{cases}
    \log_2{\frac{\const{M}}{1}}-\Bigl(\log_2{\frac{2}{1}}\Bigr)(\const{D}-1)
    \\ & \hspace{-3cm} \textnormal{if } 1\leq\const{D}\leq 2,
    \\
    \log_2{\frac{\const{M}}{2}}-\Bigl(\log_2{\frac{3}{2}}\Bigr)(\const{D}-2)
    \\ & \hspace{-3cm} \textnormal{if } 2<\const{D}\leq 3,
    \\
    & \hspace{-3cm} \vdots
    \\
    \log_2{\frac{\const{M}}{\const{M}-1}}-\Bigl(\log_2{\frac{\const{M}}{\const{M}-1}}\Bigr)(\const{D}-(\const{M}-1))
    \\
    & \hspace{-3cm} \textnormal{if } \const{M}-1<\const{D}\leq\const{M}.
  \end{cases}
\end{IEEEeqnarray*}
See also the pictorial illustration in Fig.~\ref{Fig:rho_D}. For instance, the red line $\log_2{\frac{\const{M}}{w-1}}-\bigl(\log_2{\frac{w}{w-1}}\bigr)(\const{D}-(w-1))$ going through the points $(w-1,\log_2{\frac{\const{M}}{w-1}})$ and $(w,\log_2{\frac{\const{M}}{w}})$ has the largest function values over the interval $[w-1,w]$.

Since the leakage-download function of an arbitrary WPIR scheme is bounded from below by $\rho_\mathsf{LB}^{(\mathsf{MI})}(\const{D})$, and it can be shown that the pair $\bigl(\bar{\varrho},\const{D}^{(\mathsf{MI})}(\bar{\varrho})\bigr)$ of Theorem~\ref{thm:capacity_SWPIR_MI} lies on $\rho_\mathsf{LB}^{(\mathsf{MI})}(\const{D})$ (details are omitted for brevity), this completes the converse proof.


\section{Converse of Theorem~\ref{thm:capacity_SWPIR_MaxL}}
\label{sec:converse_Thm2}

Following an argumentation similar to the first part of the converse proof of Theorem~\ref{thm:capacity_SWPIR_MI}, since the MaxL metric also satisfies the data processing inequality from the first item of Lemma~\ref{lem:convex_MI-MaxL}, the leakage-download function for the MaxL privacy metric can be bounded from below by
\begin{IEEEeqnarray}{c}
  \min_{p(\vect{u}|m)\colon\E{L(M,\vect{U})}\leq\const{D}}\ML{M}{\vect{U}}.\label{eq:leakage-download_MaxL}
\end{IEEEeqnarray}

This by itself is not a convex minimization problem. In this proof, we derive a lower bound to \eqref{eq:leakage-download_MaxL} directly. To make the problem tractable, we use the fact that $2^{\ML{M}{\vect{U}}}$ is convex in $P_{\vect{U}|M}$ (see 2) in Lemma~\ref{lem:convex_MI-MaxL}).


We know from \eqref{eq:expression_MaxL} that maximizing the objective function $\ML{M}{\vect{U}}$ is equivalent to maximizing the function
\begin{IEEEeqnarray*}{c}
  2^{\ML{M}{\vect{U}}}=\sum_{\vect{u}\in\{0:1\}^{\const{M}}}\max_{m\in[\const{M}]} p(\vect{u}|m).
\end{IEEEeqnarray*}
Moreover, from \eqref{eq:definition_PU} we know that
\begin{IEEEeqnarray*}{c}
  p(\vect{u}|m)=0,\qquad\textnormal{if }\vect{u}\in\{0,1\}^{\const{M}},\,m\notin\chi(\vect{u}).
\end{IEEEeqnarray*}
Thus,
\eqref{eq:leakage-download_MaxL} can be re-written as the convex minimization problem
\begin{IEEEeqnarray}{rCl}
  \IEEEyesnumber\label{eq:leakage-download_MaxL-PMFs}
  \IEEEyessubnumber*
  \textnormal{minimize} & &\quad\sum_{\vect{u}\in\{0,1\}^\const{M}}\max_{m\in\Spt{\vect{u}}}p(\vect{u}|m)
  \label{eq:objective-ft_MaxL-PMFs}\\[2mm]
  \textnormal{subject to} & &\quad\sum_{\vect{u}\in\{0,1\}^\const{M},m\in[\const{M}]}P_M(m)p(\vect{u}|m)L(m,\vect{u})\leq\const{D},
  \nonumber\\*\IEEEeqnarraynumspace\label{eq:download-constraint_MaxL-PMFs}\\[1mm]
  & &\quad\sum_{\vect{u}\in\{0,1\}^{\const{M}}}p(\vect{u}|m)=1,\quad\forall\,m\in[\const{M}].
  \IEEEeqnarraynumspace\label{eq:SumToOne_MaxL-PMFs}  
\end{IEEEeqnarray}
\begin{table*}[t!]
  \centering
  \caption{The equivalent LP problem of~\eqref{eq:leakage-download_MaxL-y} with the objective function $\vect{c}\trans{[\vect{y},\vect{z}]} = -\sum_{w=2}^{\const{M}}\bigl(1-\frac{1}{w}\bigr)y_w$, where the first row indicates the variables $[\vect{y},\vect{z}]=[y_2,\ldots,y_\const{M},z_1,z_2]$, the second row represents the coefficients $\vect{c}$, and the third and fourth rows are obtained from the constraints $\mat{A}\trans{[\vect{y},\vect{z}]}=\trans{\vect{b}}$.}
  \vskip -5.0ex %
  \label{tab:simplex_leakage-download_MaxL-y}
  \begin{IEEEeqnarray*}{c}
    \begin{IEEEeqnarraybox}[
      \IEEEeqnarraystrutmode
      \IEEEeqnarraystrutsizeadd{4pt}{5pt}]{v/c/v/c/v/c/v/c/v/c/v/c/v/c/v/c/v/c/v/c/v/c/v}
      \IEEEeqnarrayrulerow\\
      & && y_2 && y_3 && \cdots && y_{w-1} && y_{w} && \cdots && y_{\const{M}} && z_1 && z_2 && \vect{b} &
      \\\hline
      & \vect{c} && -\biggl[1-\frac{1}{2}\biggr]&& -\biggl[1-\frac{1}{3}\biggr] &&\cdots && -\biggl[1-\frac{1}{w-1}\biggr] && -\biggl[1-\frac{1}{w}\biggr] && \cdots && -\biggl[1-\frac{1}{\const{M}}\biggr] && 0 && 0 &&  &
      \\*\hline
      &\multirow{2}{*}{$\mat{A}$} && 1  && 1 && \cdots && 1      && 1      && \cdots && 1             && 1   &&  0  &&\const{M} &
      \\*
      & && 1  && 2 && \cdots && w-2    && w-1    && \cdots && \const{M}-1   && 0   &&  1  && \const{M}(\const{D}-1) &
      \\*\IEEEeqnarrayrulerow
    \end{IEEEeqnarraybox}
  \end{IEEEeqnarray*}
\end{table*}

Furthermore, using the fact that
\begin{IEEEeqnarray*}{c}
  \sum_{m\in\Spt{\vect{u}}}p(\vect{u}|m)\leq\Hwt{\vect{u}}\max_{m\in\Spt{\vect{u}}}p(\vect{u}|m),\quad\forall\,\vect{u}\in\{0,1\}^{\const{M}},
\end{IEEEeqnarray*}
and
\begin{IEEEeqnarray}{c}
  p(\vect{1}_{\{m\}}|m)=1-\sum_{\substack{\vect{u}\colon m\in\chi(\vect{u})\\\Hwt{\vect{u}}>1}}p(\vect{u}|m),\quad\forall\,m\in[\const{M}],
  \IEEEeqnarraynumspace\label{eq:mth-PMF}
\end{IEEEeqnarray}
the objective function \eqref{eq:objective-ft_MaxL-PMFs} becomes\footnote{In the following, the ranges of the summations and also the explicit summation variable are sometimes omitted as they are clear from the context.}
\begin{IEEEeqnarray}{rCl}
  \IEEEeqnarraymulticol{3}{l}{%
    \sum_{\vect{u}}\max_{m\in\Spt{\vect{u}}}p(\vect{u}|m)}
  \nonumber\\*[1mm]
  & = &
  \!\!\!\sum_{\vect{u}\colon \Hwt{\vect{u}}=1}p(\vect{u}|\Spt{\vect{u}})
  +\sum_{\vect{u}\colon \Hwt{\vect{u}}>1}\max_{m\in\Spt{\vect{u}}}p(\vect{u}|m)
  \nonumber\\[1mm]
  & = &
  \!\!\!\sum_{m\in[\const{M}]}\Biggl[1-\!\!\!\sum_{\substack{\vect{u}\colon m\in\chi(\vect{u})\\\Hwt{\vect{u}}>1}}p(\vect{u}|m)\Biggr]
  +\!\!\!\!\!\sum_{\vect{u}\colon \Hwt{\vect{u}}>1}\max_{m\in\Spt{\vect{u}}}p(\vect{u}|m)
  \nonumber\\[1mm]
  & \geq &
  \const{M}-\!\!\!\sum_{m\in[\const{M}]}\sum_{\substack{\vect{u}\colon m\in\chi(\vect{u})\\\Hwt{\vect{u}}>1}}p(\vect{u}|m)
  +\!\!\!\!\!\sum_{\vect{u}\colon\Hwt{\vect{u}}>1}\sum_{m\in\Spt{\vect{u}}}\frac{p(\vect{u}|m)}{\Hwt{\vect{u}}}
  \nonumber\\
  & \stackrel{(a)}{=} &
  \const{M}-\!\!\!\sum_{\vect{u}\colon\Hwt{\vect{u}}>1}\sum_{m\in\Spt{\vect{u}}}\left(1-\frac{1}{\Hwt{\vect{u}}}\right)p(\vect{u}|m),
  \IEEEeqnarraynumspace\label{eq:use_interchang-summations}\nonumber
\end{IEEEeqnarray}
where $(a)$ holds since by expanding the double summation, one can see that
\begin{IEEEeqnarray*}{c}
  \sum_{m\in[\const{M}]}\sum_{\vect{u}\colon m\in\chi(\vect{u})\atop\Hwt{\vect{u}}>1}p(\vect{u}|m)=\sum_{\vect{u}\colon\Hwt{\vect{u}}>1}\sum_{m\in\Spt{\vect{u}}}p(\vect{u}|m).
\end{IEEEeqnarray*}

Similarly, by substituting \eqref{eq:mth-PMF} into the download cost constraint \eqref{eq:download-constraint_MaxL-PMFs}, we get
\begin{IEEEeqnarray*}{rCl}
  \IEEEeqnarraymulticol{3}{l}{%
    \sum_{\vect{u}}\sum_{m\in\Spt{\vect{u}}}p(\vect{u}|m) L(m,\vect{u})
  }\nonumber\\*
  & = &\sum_{\Hwt{\vect{u}}=1}\sum_{m\in\Spt{\vect{u}}}p(\vect{u}|\Spt{\vect{u}})\cdot 1
  \nonumber\\
  && \>+\sum_{\Hwt{\vect{u}}>1}\sum_{m\in\Spt{\vect{u}}}p(\vect{u}|m)\Hwt{\vect{u}}
  \nonumber\\[1mm]
  & = &\const{M}-\sum_{\Hwt{\vect{u}}>1}\sum_{m\in\Spt{\vect{u}}}p(\vect{u}|m)
  \nonumber\\
  && \>+\sum_{\Hwt{\vect{u}}>1}\Hwt{\vect{u}}\sum_{m\in\Spt{\vect{u}}}p(\vect{u}|m)
  \IEEEeqnarraynumspace\\
  & = &\const{M}+\sum_{\Hwt{\vect{u}}>1}(\Hwt{\vect{u}}-1)\left[\sum_{m\in\Spt{\vect{u}}}p(\vect{u}|m)\right]\leq\const{M}\const{D}.\IEEEeqnarraynumspace
\end{IEEEeqnarray*}

Next, define $y_w\eqdef\sum_{\vect{u}\colon\Hwt{\vect{u}}=w}\sum_{m\in\Spt{\vect{u}}}p(\vect{u}|m)$ for $w\in[2:\const{M}]$. Because
\begin{IEEEeqnarray*}{rCl}
  \IEEEeqnarraymulticol{3}{l}{%
    \sum_{\vect{u}\colon\Hwt{\vect{u}}>1}(\Hwt{\vect{u}}-1)\sum_{m\in\Spt{\vect{u}}}p(\vect{u}|m)}
  \nonumber\\*\hspace*{2.0cm}%
  & = &\sum_{w=2}^{\const{M}}(w-1)\sum_{\vect{u}\colon\Hwt{\vect{u}}=w}\sum_{m\in\Spt{\vect{u}}}p(\vect{u}|w),
\end{IEEEeqnarray*}
it can be shown that a lower bound to \eqref{eq:leakage-download_MaxL-PMFs} can be computed from the linear programming (LP) formulation 
\begin{IEEEeqnarray}{rCl}
  \IEEEyesnumber\label{eq:leakage-download_MaxL-y}
  \IEEEyessubnumber*
  \textnormal{minimize} & &\quad\rho_{\mathsf{LB}}^{(\mathsf{MaxL})}(\const{D})\eqdef\const{M}-\sum_{w=2}^{\const{M}}\left[1-\frac{1}{w}\right]y_w
  \label{eq:objective-ft_MaxL-y}\\[1mm]
  \textnormal{subject to}& &\quad \sum_{w=2}^{\const{M}} y_w\leq \const{M},
  \IEEEeqnarraynumspace\label{eq:PMFs_Bounds_MaxL-y}
  \\[1mm]
  & &\quad\sum_{w=2}^{\const{M}}(w-1)\cdot y_w\leq\const{M}(\const{D}-1),
  \label{eq:PMFs_download-constraint_MaxL-y}
\end{IEEEeqnarray}
with variables $y_w$, $w\in[2:\const{M}]$.

We convert the inequalities in the constraints \eqref{eq:PMFs_Bounds_MaxL-y} and \eqref{eq:PMFs_download-constraint_MaxL-y} to equalities by introducing variables $z_1$ and $z_2$. Thus, we have the constraints as
\begin{align*}
	&\sum_{w=2}^{\const{M}} y_w+z_1= \const{M},\\
	&\sum_{w=2}^{\const{M}}(w-1)\cdot y_w+z_2=\const{M}(\const{D}-1).
\end{align*}
Now, define $c_w\eqdef -\bigl(1-\frac{1}{w}\bigr)$, for $w\in[2:\const{M}]$, and let $c_{\const{M}+1}=c_{\const{M}+2}= 0$. Then, the objective function of \eqref{eq:leakage-download_MaxL-y} can be written in matrix form as $\const{M}+\vect{c}\trans{[y_2,\ldots,y_\const{M},z_1,z_2]}=\const{M}-\sum_{w=2}^{\const{M}}\bigl(1-\frac{1}{w}\bigr)y_w$, with $\bm c = (c_2,\ldots,c_{\const{M}+2})$, and the constraints as $\mat{A} \trans{[\vect{y},\vect{z}]}=\trans{\vect{b}}$, where
\begin{IEEEeqnarray*}{rCl}
  \mat{A}=
  \begin{pmatrix}
    1 & 1 &\cdots & 1 & 1 & 0 
    \\
    1 & 2 &\cdots &\const{M}-1 & 0 & 1
  \end{pmatrix},\quad
  \trans{\vect{b}}=
  \begin{pmatrix}
    \const{M}
    \\
    \const{M}(\const{D}-1)
  \end{pmatrix},
\end{IEEEeqnarray*}
  $\vect{y} = (y_2,\ldots,y_\const{M})$, and $\vect{z} = (z_1,z_2)$.  
The equivalent LP problem of~\eqref{eq:leakage-download_MaxL-y} (without the constant $\const{M}$) is shown in Table~\ref{tab:simplex_leakage-download_MaxL-y}.

Consider the standard LP problem of minimizing $\vect{c}\trans{\vect{x}}$ subject to $\mat{A}\trans{\vect{x}}=\trans{\vect{b}}$ and  $\vect{x}\geq \bm 0$, where $\mat{A}$ is an arbitrary $m\times n$ matrix. A \emph{basic solution} is any  solution where a subset of $n-m$ variables are zero.  The $m$ nonzero variables of a basic solution are referred to as the basic variables, while the remaining variables are known as the nonbasic variables. Let $\mat{A}_{\set{B}}$ denote the submatrix of $\mat{A}$ consisting of the $m$ columns of $\mat{A}$ corresponding to the basic variables, and by $\mat{A}_{\set{N}}$ the submatrix consisting of the remaining $n-m$ columns  corresponding to the nonbasic variables. 
Denote by $\vect{x}_\set{B}$ the basic variables of a basic solution $\vect{x}$, and let $\vect{c}_\set{B}$ and $\vect{c}_{\set{N}}$ be the subvectors of $\vect{c}$ that correspond to the basic and nonbasic variables, respectively. The following proposition provides a sufficient condition for a basic solution to be optimal for an LP problem.\footnote{The proof of Proposition~\ref{prop:sufficient_LP} can be found in the LP literature, see, e.g.,~\cite[pp.~43--44]{LuenbergerYe16_1}.}
\begin{proposition}[{\cite[p.~44]{LuenbergerYe16_1}}]
  \label{prop:sufficient_LP}
  If there exists a basic solution such that the \emph{relative cost vector} for nonbasic variables, defined as $\vect{c}_\set{N}-\vect{c}_{\set{B}}\inv{\mat{A}_\set{B}}\mat{A}_\set{N}$, is nonnegative, then the basic solution is optimal.
\end{proposition}

From the capacity-achieving scheme proposed in Section~\ref{sec:achievability}, one can show that
\begin{IEEEeqnarray*}{c}
  p^{\ast}(\vect{u}|m)=
  \begin{cases}
    \frac{w-\const{D}}{\binom{\const{M}-1}{w-2}} & \textnormal{if }\Hwt{\vect{u}}=w-1,
    \\[2mm]      
    \frac{\const{D}-(w-1)}{\binom{\const{M}-1}{w-1}} & \textnormal{if }\Hwt{\vect{u}}=w,
    \\[2mm]
    0 & \textnormal{otherwise},
  \end{cases}
\end{IEEEeqnarray*}
for $w-1\leq\const{D}\leq w$, $w\in[2:\const{M}]$, where the superscript $\ast$ indicates that the corresponding quantity is for the particular scheme from  Section~\ref{sec:achievability}. Then, since $y^\ast_w=\sum_{\vect{u}\colon\Hwt{\vect{u}}=w}\sum_{m\in\Spt{\vect{u}}}p^\ast(\vect{u}|m)$ for $w\in[2:\const{M}]$, we have, for $1\leq \const{D}\leq 2$,
\begin{IEEEeqnarray*}{l}
  \begin{IEEEeqnarraybox}[\mystrut][c]{rCl}
    y^\ast_2& = &\sum_{\vect{u}\colon\Hwt{\vect{u}}=2}\sum_{m\in\Spt{\vect{u}}}\frac{\const{D}-1}{\const{M}-1}=\const{M}(\const{D}-1),
    \\
    y^\ast_{w'}& = &0,\quad w'\in[3:\const{M}], 
    \\
    z_1^*& = &\const{M}(2-\const{D}),\quad z_2^\ast=0,
  \end{IEEEeqnarraybox}
 \end{IEEEeqnarray*}
 and 
  \begin{IEEEeqnarray*}{l}
  \begin{IEEEeqnarraybox}[\mystrut][c]{rCl}
    y^\ast_{w-1}& = &\binom{\const{M}}{w-1}(w-1)\frac{w-\const{D}}{\binom{\const{M}-1}{w-2}}=\const{M}(w-\const{D}),
    \\[2mm]
    y^\ast_w& = &\binom{\const{M}}{w}w\frac{\const{D}-(w-1)}{\binom{\const{M}-1}{w-1}}=\const{M}\bigl(\const{D}-(w-1)\bigr),
    \\
    y^\ast_{w'}& = &0,\quad w'\in[3:\const{M}] \setminus \{ w-1,w\}, 
    \\
    z_1^\ast& = &0,\quad z_2^\ast = 0,
  \end{IEEEeqnarraybox}
\end{IEEEeqnarray*}  
for $w-1\leq \const{D}\leq w$, $w\in[3:\const{M}]$. 
To prove the converse part, we use Proposition~\ref{prop:sufficient_LP} to verify the optimality of $y^\ast_w$, $w\in[2:\const{M}]$, and $z^\ast_1$, $z^\ast_2$, for~\eqref{eq:leakage-download_MaxL-y}, by considering two special cases as follows. 


\begin{description}[leftmargin=0cm]
\item[Case 1.] $1\leq\const{D}\leq 2$: In this case, $y_2$ and $z_1$ are the basic variables. Since
  \begin{IEEEeqnarray*}{rCl}
    \IEEEeqnarraymulticol{3}{l}{%
      c_w-\vect{c}_\set{B}\inv{\mat{A}_\set{B}}\mat{A}_{\{y_w\}}
    }\nonumber\\*\quad%
    & = &-\biggl[1-\frac{1}{w}\biggr]-\Bigl(-\frac{1}{2},0\Bigr)\inv{
      \begin{pmatrix}
        1 & 1
        \\
        1 & 0
      \end{pmatrix}}
    \begin{pmatrix}
      1
      \\
      w-1
    \end{pmatrix}
    \\[1mm]
    & = &\frac{w-1}{2}-\frac{w-1}{w}> 0,
  \end{IEEEeqnarray*}
  for $w\in[3:\const{M}]$, and $0-\bigl(-\frac{1}{2},0\bigr)\inv{
      \bigl(\begin{smallmatrix}
        1 & 1
        \\
        1 & 0
      \end{smallmatrix}\bigr)}
    \bigl(\begin{smallmatrix}
      0
      \\
      1
    \end{smallmatrix}\bigr)=\frac{1}{2}>0$ for $z_2$, it follows that the basic solution is optimal for $\const{M}\geq 2$.

\item[Case 2.] $w-1<\const{D}\leq w$, $w\in[3:\const{M}]$: In this case, $y_{w-1}$ and $y_w$ are the basic variables. We can see that for $w'\in[3:\const{M}] \setminus \{w-1,w\}$, the relative cost coefficient for $y_{w'}$ is  
  \begin{IEEEeqnarray*}{rCl}
    \IEEEeqnarraymulticol{3}{l}{%
      c_{w'}-\vect{c}_\set{B}\inv{\mat{A}_\set{B}}\mat{A}_{\{y_w'\}}}\nonumber\\*\quad%
    & = &
    -\biggl[1-\frac{1}{w'}\biggr]-\biggl(-\Bigl[1-\frac{1}{w-1}\Bigr],-\Bigl[1-\frac{1}{w}\Bigr]\biggr)
    \nonumber\\[1mm]
    && \>\cdot\inv{
      \begin{pmatrix}
        1 & 1
        \\
        w-2 & w-1
      \end{pmatrix}}
    \begin{pmatrix}
      1
      \\
      w'-1
    \end{pmatrix}
    \\[1mm]
    & = &\frac{1}{w'}-\frac{w-w'}{w-1}+\frac{w-w'-1}{w}.
  \end{IEEEeqnarray*}
  By performing some simple calculations, we get
  \begin{IEEEeqnarray*}{rCl}
    \IEEEeqnarraymulticol{3}{l}{%
      \frac{1}{w'}-\frac{w-w'}{w-1}+\frac{w-w'-1}{w}}\nonumber\\*\hspace*{3.0cm}%
    & = &(w-w')\biggl[\frac{w-(w'+1)}{w\cdot w'(w-1)}\biggr]>0,\IEEEeqnarraynumspace
  \end{IEEEeqnarray*}
  either for $w'>w-1$ or $w'<w$. Moreover, it is easy to see that the relative cost coefficients for $z_1$ and $z_2$ are
    $\frac{w-2}{w}>0$
  and $\frac{1}{w-1}-\frac{1}{w} >0$, respectively, which implies that the given basic solution is optimal.
\end{description}

\section{Conclusion}
\label{sec:conclusion}
We considered relaxing the perfect privacy condition in single-server PIR and presented a scheme for the studied weakly-private scenario, referred to as WPIR. In doing so, we showed that one can trade off privacy to gain in terms of download cost. Furthermore, we characterized the information leaked using two different metrics: MI and MaxL. The latter is known to be a more robust metric to measure information leakage. Finally, we derived the single-server WPIR capacity for both the MI and MaxL metrics, and showed that the proposed protocol is capacity-achieving. As a final note, we drew the connection between WPIR and rate-distortion theory.

An interesting direction for future work is the derivation of fundamental bounds on other performance metrics like the upload cost and the access complexity for the single server scenario.

\section*{Acknowledgment}

The authors would like to thank the three anonymous reviewers and the Guest Editor Prof.~Lalitha Sankar for their valuable and insightful comments.

\IEEEtriggeratref{32}

\end{document}